\documentclass[12pt,journal,onecolumn]{IEEEtran}
\usepackage{amsfonts}
\usepackage{bbm}
\usepackage{mathrsfs}
\usepackage{graphicx}
\ifCLASSINFOpdf
\else
\fi
\usepackage{amsmath}
\usepackage{amssymb}
\usepackage{amsbsy}
\usepackage{amstext}
\usepackage{amsopn}
\usepackage{amsthm}
\usepackage{amssymb}
\usepackage{eufrak}
\usepackage{cite}

\usepackage{fancybox}
\usepackage{dashbox}
\usepackage{threeparttable}

\newtheorem{theorem}{Theorem}
\newtheorem{definition}{Definition}

\newtheorem{proposition}{Proposition}
\renewcommand{\IEEEQED}{\IEEEQEDopen}
\newcommand{\tqbinom}[2]{\genfrac{[}{]}{0pt}{}{#1}{#2}}

\newenvironment{ventry}[1]%
{\begin{list}{}{%
\settowidth{\labelwidth}{\textsf{#1}}%
\setlength{\labelsep}{5mm}
\setlength{\leftmargin}{\labelwidth+\labelsep}}}%
{\end{list}}
\begin{document}
\title{Group-Decodable Space-Time Block Codes\\ with Code Rate $>$ 1}
\author{Tian Peng Ren,~Yong Liang Guan,~Chau Yuen,~Erry Gunawan and~Er Yang
Zhang
}

\maketitle

\begin{abstract}
High-rate space-time block codes (STBC with code rate $>1$) in multi-input multi-output (MIMO) systems are able to provide both spatial multiplexing gain and diversity gain, but have high maximum likelihood (ML) decoding complexity. Since group-decodable (quasi-orthogonal) code structure can reduce the decoding complexity, we present in this paper systematic methods to construct group-decodable high-rate STBC with full symbol-wise diversity gain for arbitrary transmit antenna number and code length. We show that the proposed group-decodable STBC can achieve high code rate that increases almost linearly with the transmit antenna number, and the slope of this near-linear dependence increases with the code length. Comparisons with existing low-rate and high-rate codes (such as orthogonal STBC and algebraic STBC) are conducted to show the decoding complexity reduction and good code performance achieved by the proposed codes.
\end{abstract}
\IEEEpeerreviewmaketitle

\section{Introduction}

\indent Space-time codes (STC) in multi-input multi-output (MIMO) systems have been extensively studied for their
ability to provide transmit diversity gain and spatial multiplexing gain \cite{Zheng}. Space-time
trellis codes (STTC) \cite{Tarokh_stc} and space-time block codes (STBC) \cite{Alamouti,Tarokh_ostbc,Ganesan_ostbc,Lu_ostbc,Tirkkonen,Jafarkhani,Yuen_book,Dao_4gp,Rajan_multigp,xia_linear} are able to provide diversity gain and have code rate limited by 1. On the other hand, Bell Labs layered space-time (BLAST) system \cite{Foschini}, high-rate linear dispersion (LD) codes \cite{Hassibi}, Golden code \cite{Belfiore}, perfect codes \cite{Oggier}, PS-SR code\cite{Rajan4x2}, etc., have code rate $>1$ and are able to provide multiplexing gain (the latter four have diversity gain too).

To achieve higher code rates with low joint-decoding complexity, many STBC with code rate $\leq1$ have been designed to be group-decodable (quasi-orthogonal) \cite{Tirkkonen,Jafarkhani,Yuen_book,Dao_4gp,Rajan_multigp}. In contrast, there were much fewer designs of group-decodable STBC with code rate $>1$ (high-rate STBC). In \cite{Yuen_2gp}, square 2-group-decodable STBC of code rate 1.25 for 4 transmit antennas were obtained by computer search; In \cite{Rajan_2gp}, 2-group-decodable STBC of code rate $2^{m-2}+\frac{1}{2^m}$ for $2^m$ ($m\geq 2$) transmit antennas were constructed. In \cite{Ren_de}, it was also shown that the group-decodable code structure is beneficial to diversity-embedded (DE) space-time codes as it avoids interference between the different diversity layers in the DE codes and helps to guarantee the designed diversity levels.

In this paper, group-decodable high-rate STBC with arbitrary number of transmit antennas and code lengths is considered, then systematic methods to construct them with full symbol-wise diversity are presented. Their maximum achievable code rate and decoding complexity are analyzed. Specific code examples are constructed and simulated.

The rest of this paper is organized as follows. High-rate STBC with code rate $>1$ will be abbreviated as STBC. In Section II, the system model is described and group-decodable STBC is
defined. Unbalanced 2-group-decodable STBC and balanced
2-group-decodable STBC are constructed systematically in Section III
and Section IV, respectively. Comparisons of the decoding complexity
and BER performance are shown in Section V. Finally, this paper is
concluded in Section VI.

In this paper, bold lower case and upper case letters
denote vectors and matrices (sets), respectively; $\mathbb{R}$ and
$\mathbb{C}$ denote the real and the complex number field, respectively;
$(\cdot)^R$ and $(\cdot)^I$ stand for the real and the imaginary part
of a complex vector or matrix, respectively; $[\cdot]^*$,
$[\cdot]^T$, $[\cdot]^H$ and $\| \cdot \|$ denote the complex
conjugate, the transpose, the complex conjugate transpose and the
Frobenius norm of a matrix, respectively; $dim(\cdot)$ and
$rank(\cdot)$ represent dimension of a vector/matrix space and rank
of a matrix, respectively; $\textbf{I}$ denotes an identity matrix.

\section{System Model}
\subsection{Signal Model}
We consider a space-time block coding system employing
$N$ transmit antennas and $M$ receive antennas. The transmitted
signal sequences are partitioned into independent time blocks for
transmission over $T$ symbol durations using STBC matrix $\textbf{X}$
of size $T\times N$. Following the signal model in \cite{Hassibi},
$\textbf{X}$ can be denoted as:
\begin{equation}\label{LSTBC}
\textbf{X}_{T\times N}=\sum^L_{l=1}{s_{l}\textbf{C}_{l}}
\end{equation}
where $s_l\in \mathbb{R}$ are real valued symbols representing the real and imaginary components of complex constellation symbols, $\textbf{C}_l\in \mathbb{C}^{T\times N}$ are called dispersion
matrices. Thus, the code rate is $\frac{L}{2T}$
considering complex symbol transmission. The average energy of the
code matrix is constrained to $\mathcal {E}_\textbf{X}=\mathbb{E}\|\textbf{X}\|^2=T$.

The received signals $\tilde{r}_{tm}$ of the $m$th receive
antenna at time $t$ can be arranged in a $T\times M$ matrix
$ [\tilde{\textbf{r}}_1~\tilde{\textbf{r}}_2~\cdots~\tilde{\textbf{r}}_M]=
[\tilde{r}_{tm}]$. Thus, the transmit-receive signal relationship can be presented as:
\begin{equation}\label{R_XH}
[\tilde{\textbf{r}}_1~\tilde{\textbf{r}}_2~\cdots~\tilde{\textbf{r}}_M]
=\sqrt{\rho}\textbf{X}\tilde{\textbf{H}}+\tilde{\textbf{Z}}
\end{equation}
where $\tilde{\textbf{H}}_{N\times M} =
[\tilde{\textbf{h}}_1~\tilde{\textbf{h}}_2~\cdots~\tilde{\textbf{h}}_M]$
is the channel matrix with independent entries $\tilde{h}_{nm}$; $\tilde{\textbf{Z}}_{T\times M}
=[\tilde{\textbf{z}}_1~\tilde{\textbf{z}}_2~\cdots$ $\tilde{\textbf{z}}_M]=[\tilde{z}_{tm}]$ is the additive noise matrix with independently, identically
distributed (i.i.d.) $\mathcal {CN}(0,1)$ entries $\tilde{z}_{tm}$; $\rho$ is the average
signal-to-noise ratio (SNR) at each receive antenna. The received
signal can also be rewritten as \cite{Hassibi}:
\begin{equation}\label{r_Hs}
\textbf{r}=\sqrt{\rho} \textbf{H}\textbf{s}+\textbf{z}
\end{equation}
where
\begin{equation*}
\begin{split}
 &\textbf{r}=\left[
\begin{array}{cccccccc}
    \tilde{\textbf{r}}^R_1\\
    \tilde{\textbf{r}}^I_1\\
    \vdots\\
    \tilde{\textbf{r}}^R_{M}\\
    \tilde{\textbf{r}}^I_{M}
\end{array}
\right],~
 \bar{\textbf{h}}=\left[
\begin{array}{cccccccc}
    \tilde{\textbf{h}}^R_{1}\\
    \tilde{\textbf{h}}^I_{1}\\
    \vdots \\
    \tilde{\textbf{h}}^R_{M}\\
    \tilde{\textbf{h}}^I_{M}
\end{array}
\right],~\textbf{s}=\left[
\begin{array}{cccccccc}
    s_1\\
    s_2\\
    \vdots\\
    s_L
\end{array}
\right],~\textbf{z}=\left[
\begin{array}{cccccccc}
    \tilde{\textbf{z}} ^R_1\\
    \tilde{\textbf{z}} ^I_1\\
    \vdots\\
    \tilde{\textbf{z}} ^R_{M}\\
    \tilde{\textbf{z}} ^I_{M}
\end{array}
\right],
\end{split}
\end{equation*}
\begin{equation*}
\begin{split}
&\textbf{H}=\left[\textbf{h}_1~\textbf{h}_2~\cdots~\textbf{h}_L\right]=\left[
\begin{array}{cccccccc}
    \mathscr{C}_1\bar{\textbf{h}} &\mathscr{C}_2\bar{\textbf{h}}  &\cdots  &\mathscr{C}_L\bar{\textbf{h}}
\end{array}
\right],
\\&
\mathscr{C}_l=\left[
\begin{array}{cccccccc}
    \mathcal {C}_l   &   \textbf{0} & \cdots & \textbf{0}\\
    \textbf{0}   &   \mathcal {C}_l & \cdots & \textbf{0}\\
    \vdots & \vdots & \ddots & \vdots \\
    \textbf{0} & \textbf{0} & \cdots & \mathcal {C}_l
\end{array}
\right]_{2TM\times 2NM},~\mathcal {C}_l=\left[
\begin{array}{cccccccc}
    \textbf{C}^R_l   &   -\textbf{C}^I_l\\
    \textbf{C}^I_l   &   \textbf{C}^R_l
\end{array}
\right]_{2T\times 2N}
\end{split}
\end{equation*}
and $l=1,2,\cdots,L$.

The maximum likelihood (ML) decoding of STBC is to find the solution
$\hat{\textbf{s}}$ so that
\begin{equation}\label{Ml_metric}
\hat{\textbf{s}}=arg\underset{\textbf{s}}{min}\|\textbf{r}-\sqrt{\rho}
\textbf{H}\textbf{s}\|^2
\end{equation}

To avoid rank deficiency at the decoder, $rank(\textbf{H})=L$ is
required, which means that $\textbf{H}$ should be \textquoteleft
\textquoteleft tall\textquoteright\textquoteright, i.e., $L\leq
2TN_r$\cite{Hassibi}\cite{Yuen_book}, which implies that the
receiver antenna number satisfies $N_r\geq \frac{L}{2T}$.

\subsection{$\Gamma$-Group-Decodable STBC}

Firstly, \emph{linear independence of
matrices} is defined as follows:

\begin{definition}\label{rank_matrix}
\emph{The matrices $\textbf{A}_1,\textbf{A}_2,\cdots,\textbf{A}_L$
are said to be linearly independent if no nontrivial linear
combination of them is equal to zero. In other words, with
$\alpha_i\in \mathbb{C}~(i=1,\cdots,L)$
\begin{equation*}
\alpha_1\textbf{A}_1+\alpha_2\textbf{A}_2+\cdots+\alpha_L\textbf{A}_L=\textbf{0}
\end{equation*}
only when $\alpha_1=\alpha_2=\cdots=\alpha_L=0$.}
\end{definition}

It is easy to show that the linear independence among
$\textbf{A}_1,\textbf{A}_2,\cdots,\textbf{A}_L$ is equivalent to the
linear independence among vectors
$\bar{\textbf{a}}_1,\bar{\textbf{a}}_2,\cdots,\bar{\textbf{a}}_L$,
where $\bar{\textbf{a}}_i=g(\textbf{A}_i)$ with $i=1,2,\cdots,L$, and $g$ is the
matrix-to-vector mapping function $\tqbinom{a~c}{b~d}\xrightarrow{g}[a~b~c~d]^T$.

The main idea of group-decodable STBC is to divide the $L$
real transmitted symbols embedded in a code matrix into several orthogonal
groups such that after linear channel matched filtering, the ML
detection metric of the transmitted symbols can be decoupled into
independent submetrics, each containing a smaller group of symbols. Assume that the transmitted symbols can be
separated into $\Gamma$ groups and each group has $L_i$ symbols,
then $\sum^\Gamma_{i=1}{L_i}=L$. Let the set of indexes of symbols
in the $i$th group be denoted as $\Theta_i$. For an STBC to be
$\Gamma$-group-decodable, two conditions should be satisfied:

(i) $\textbf{h}^T_p\textbf{h}_q=0$ where $p\in
\Theta_{i_1},~q\in \Theta_{i_2}\emph{\emph{ and }}i_1\neq i_2$;

(ii) $rank(\textbf{H}_i)=L_i$ where
$\textbf{H}_i=[\textbf{h}_{i_1}~\textbf{h}_{i_2}~\cdots~\textbf{h}_{i_{L_i}}]$,
$i_k\in \Theta_i,~k=1,2,\cdots,L_i\emph{\emph{ and }}i=1,2,\cdots,\Gamma$.

Condition (i) means that the STBC is group-decodable and
condition (ii) guarantees that no decoder of any group is rank
deficient.

To satisfy the condition (i), \emph{Yuen} et al. \cite{Yuen_2gp} have
established a necessary and sufficient condition as follows:

\begin{theorem}[Quasi-Orthogonality Constraint,~QOC]\label{qoc_th}
\emph{The necessary and sufficient condition to make $s_p$ and $s_q~(p\neq q)$ in the STBC matrix (\ref{LSTBC}) to be orthogonal (i.e., to achieve $\textbf{h}^T_p\textbf{h}_q=0$) is
\begin{equation}\label{qoc}
\textbf{C}^H_p\textbf{C}_q=-\textbf{C}^H_q\textbf{C}_p.
\end{equation}}
\end{theorem}

Regarding the condition (ii), $rank(\textbf{H}_i)=L_i$
implies that $\textbf{h}_{i_1},\textbf{h}_{i_2},\cdots,\textbf{h}_{i_{L_i}}$
should be linearly independent.

\begin{theorem}\label{th_independence}
\emph{The necessary and sufficient condition for
$\textbf{h}_{i_1},\textbf{h}_{i_2},\cdots,\textbf{h}_{i_{L_i}}$ to
be linearly independent is that
$\tqbinom{\textbf{C}^R_{i_1}}{\textbf{C}^I_{i_1}}$, $
\tqbinom{\textbf{C}^R_{i_2}}{\textbf{C}^I_{i_2}},\cdots,$ $
\tqbinom{\textbf{C}^R_{i_{L_i}}}{\textbf{C}^I_{i_{L_i}}}$ must be
linearly independent.}
\end{theorem}

The proof of Theorem \ref{th_independence} is given in Appendix \ref{proofindependence}.

From the above, a formal definition of $\Gamma$-group-decodable STBC can be presented as:

\begin{definition}\label{group_decodable_STBC}
\emph{An STBC is said to be $\Gamma$-group-decodable if}

\emph{(i)
$\textbf{C}^H_p\textbf{C}_q=-\textbf{C}^H_q\textbf{C}_p,~\forall
p\in \Theta_{i_1},\forall q\in \Theta_{i_2},~{i_1}\neq i_2$;}

\emph{(ii)
$\tqbinom{\textbf{C}^R_{i_1}}{\textbf{C}^I_{i_1}},\cdots,
\tqbinom{\textbf{C}^R_{i_k}}{\textbf{C}^I_{i_k}},\cdots,
\tqbinom{\textbf{C}^R_{i_{L_i}}}{\textbf{C}^I_{i_{L_i}}}$ are
linearly independent where $i_k \in
\Theta_i,~k=1,2,\cdots,L_i,~i=1,2,\cdots,\Gamma$.}
\end{definition}

In this paper, we focus on 2-group-decodable STBC, i.e., $\Gamma=2$. For 2-group-decodable STBC, the total transmitted symbols $L=L_1+L_2$ where $L_1$ and $L_2$ are the number of symbols in the first group and second group, respectively. We will consider two cases.
The first case is $L_1=1$ and $L_2=L-1$ (a special case of $L_1\neq L_2$), called \emph{unbalanced}
2-group-decodable STBC;
the other case is $L_1=L_2=\frac{L}{2}$, called \emph{balanced}
2-group-decodable STBC. The former will be used to construct the
latter.

\section{Unbalanced 2-Group-Decodable STBC}
\subsection{Code Construction} \label{section_un_construction}

Considering the unbalanced 2-group-decodable STBC with
$L_1=1$ and $L_2=L-1$, we have
\begin{equation} \label{draft_eq_qoc}
\textbf{C}^H_1\textbf{C}_l=-\textbf{C}^H_l\textbf{C}_1,~(l=2,3,\cdots,L).
\end{equation}

For brevity, $\textbf{C}_1$ is simplified as
$\textbf{C}=[c_{tn}]$, and $\textbf{C}_l$ with $l\in\{2,3,\cdots ,L\}$ is represented by
$\textbf{Y}=[y_{tn}]$, $t=1,2,\cdots,T\emph{\emph{ and }}n=1,2,\cdots ,N$.
\begin{definition}
\emph{Symbol-wise diversity is denoted as the minimum rank of the dispersion matrices in an STBC\cite{Tirkkonen_symbolwisediversity}\cite{Hassibi}.}
\end{definition}

To achieve full symbol-wise diversity gain, $\textbf{C}$ is required to be full rank, i.e., $rank(\textbf{C})=\min(T,N)$. Then, (\ref{draft_eq_qoc}) can be written as
\begin{equation}\label{eq_draft}
\textbf{C}^H\textbf{Y}+\textbf{Y}^H\textbf{C}=\textbf{0}.
\end{equation}

It is easy to show that (\ref{eq_draft}) can be converted
into scalar equations as:
\begin{equation}\label{eq_dis}
\begin{split}
&\sum^{T}_{t=1}{c^R_{tn}y^R_{tn}+c^I_{tn}y^I_{tn}}=0\\
&\sum^{T}_{t=1}{c^R_{tn}y^R_{ti}+c^I_{tn}y^I_{ti}+c^R_{ti}y^R_{tn}+c^I_{ti}y^I_{tn}}=0\\
&\sum^{T}_{t=1}{c^I_{tn}y^R_{ti}-c^R_{tn}y^I_{ti}-c^I_{ti}y^R_{tn}+c^R_{ti}y^I_{tn}}=0
\end{split}
\end{equation}
where $n=1,2,\cdots, N$, $i=n+1,\cdots,N$. In turn, (\ref{eq_dis})
can be rewritten in matrix form as:
\begin{equation}\label{eq_matrix}
\mathcal {C}\bar{\textbf{y}}=\textbf{0}
\end{equation}
where $\mathcal {C}=f(\textbf{C})$ of size $N^2\times 2TN$ and
$\bar{\textbf{y}}=g(\tqbinom{\textbf{Y}^R}{\textbf{Y}^I})$ of size
$2TN\times 1$ with mapping functions $f$ and $g$ given in (\ref{Candy}),
$\mathbbm{c}_n=[c^R_{1n}~c^I_{1n}~c^R_{2n}~c^I_{2n}~\cdots~c^R_{Tn}~c^I_{Tn}]$ and $\mathbbm{c}'_n=[c^I_{1n}~-c^R_{1n}~c^I_{2n}~$
$-c^R_{2n}~\cdots~c^I_{Tn}~-c^R_{Tn}]$.

\begin{equation} \label{Candy}
\mathcal {C}=f(\textbf{C})=\left[
\begin{array}{ccccccccccccccccccccccccccc}
    \mathbbm{c}_1     &\textbf{0}   &\cdots &\cdots  &\cdots &\textbf{0} \\
    \textbf{0}       &\mathbbm{c}_2   &\textbf{0} &\cdots  &\cdots &\textbf{0} \\
    \vdots &\vdots  &\ddots    &\ddots &\ddots &\vdots  \\
    \textbf{0}        &\cdots &\cdots &\cdots &\textbf{0} &\mathbbm{c}_N\\
    \mathbbm{c}_2     &\mathbbm{c}_1    &\textbf{0}   &\textbf{0} &\cdots  &\textbf{0} \\
    \mathbbm{c}'_2    &-\mathbbm{c}'_1   &\textbf{0}  &\textbf{0} &\cdots  &\textbf{0} \\
    \mathbbm{c}_3     &\textbf{0}   &\mathbbm{c}_1    &\textbf{0} &\cdots  &\textbf{0} \\
    \mathbbm{c}'_3    &\textbf{0}   &-\mathbbm{c}'_1  &\textbf{0} &\cdots  &\textbf{0} \\
    \vdots &\vdots  &\ddots    &\ddots &\ddots &\vdots  \\
    \mathbbm{c}_N       &\textbf{0}   &\cdots &\cdots  &\textbf{0} &\mathbbm{c}_1  \\
    \mathbbm{c}'_N      &\textbf{0}   &\cdots &\cdots  &\textbf{0} &-\mathbbm{c}'_1 \\
    \textbf{0}      &\mathbbm{c}_3     &\mathbbm{c}_2    &\textbf{0}   &\cdots  &\textbf{0} \\
    \textbf{0}      &\mathbbm{c}'_3    &-\mathbbm{c}'_2   &\textbf{0}   &\cdots  &\textbf{0} \\
    \vdots &\vdots  &\vdots &\ddots    &\ddots &\vdots  \\
    \textbf{0}      &\mathbbm{c}_N       &\textbf{0}   &\cdots  &\textbf{0} &\mathbbm{c}_2   \\
    \textbf{0}      &\mathbbm{c}'_N     &\textbf{0}   &\cdots  &\textbf{0}  &-\mathbbm{c}'_2 \\
    \vdots &\vdots  &\vdots &\ddots    &\ddots &\vdots  \\
    \textbf{0}         &\cdots &\cdots  &\textbf{0} &\mathbbm{c}_{N}    &\mathbbm{c}_{N-1}   \\
    \textbf{0}        &\cdots &\cdots  &\textbf{0}  &\mathbbm{c}'_{N} &-\mathbbm{c}'_{N-1} \\
\end{array}
\right],~~\bar{\textbf{y}}=g(\tqbinom{\textbf{Y}^R}{\textbf{Y}^I})=\left[
\begin{array}{cccccc}
    y^R_{11} \\y^I_{11}\\\vdots\\
y^R_{T1}\\y^I_{T1}\\y^R_{12} \\y^I_{12}\\\vdots
\\y^R_{T2}\\y^I_{T2}\\\vdots \\y^R_{1N} \\y^I_{1N}\\\vdots \\y^R_{TN}\\y^I_{TN}
\end{array}
\right]
\end{equation}

As $\bar{\textbf{y}}$ is of size $2TN\times 1$, the solution
space of (\ref{eq_matrix}), $\{\bar{\textbf{y}}\}$, is of dimension
$2TN-rank(\mathcal {C})$. Let
$\bar{\textbf{y}}_1,\bar{\textbf{y}}_2,\cdots,$
$\bar{\textbf{y}}_{2TN-rank(\mathcal {C})}$ be the basis of
$\{\bar{\textbf{y}}\}$, which are linearly independent. Denoting
$g^{-1}$ as the inverse function of $g$ in (\ref{Candy}), linearly
independent matrices
$\tqbinom{\textbf{Y}^R_1}{\textbf{Y}^I_1},$
$\tqbinom{\textbf{Y}^R_2}{\textbf{Y}^I_2}, \cdots,
\tqbinom{\textbf{Y}^R_{2TN-rank(\mathcal
{C})}}{\textbf{Y}^I_{2TN-rank(\mathcal {C})}}$ can be obtained as $\tqbinom{\textbf{Y}^R_i}{\textbf{Y}^I_i}=g^{-1}(\bar{\textbf{y}}_i)$
with $i=1,2,\cdots,2TN-rank(\mathcal {C})$. From Definition \ref{group_decodable_STBC}, if $\textbf{Y}_1,\textbf{Y}_2,\cdots,\textbf{Y}_{2TN-rank(\mathcal
{C})}$ and $\textbf{C}$ in (\ref{eq_draft}) are used as the dispersion matrices, the resultant STBC will be an
unbalanced 2-group-decodable STBC of code rate $\frac{2TN-rank(\mathcal {C})+1}{2T}$ with 1 real symbol in the first
group and $2TN-rank(\mathcal {C})$ real symbols in the second group.

From the discussions above, we can summarize the systematic
construction of unbalanced 2-group-decodable STBC as follows:

Step 1: Pick a $T\times N$ matrix $\textbf{C}$ with full rank as
the dispersion matrix $\textbf{C}_1$ in the first group;

Step 2: Based on the matrix $\textbf{C}$, obtain the matrix
$\mathcal {C}=f(\textbf{C})$ following equation (\ref{Candy});

Step 3: Based on the matrix $\mathcal {C}$, solve equation (\ref{eq_matrix}) and obtain its
solution space represented as $\{\bar{\textbf{y}}_1,\bar{\textbf{y}}_2,$ $\cdots,\bar{\textbf{y}}_{2TN-rank(\mathcal
{C})}\}$ subject to the condition that all $\bar{\textbf{y}}_i$ ($i=1,\cdots,2TN-rank(\mathcal {C})$) lead to full-rank dispersion matrices in Step 4;

Step 4: Using the vector-to-matrix mapping function
$g^{-1}$ (inverse function of $g$ in (\ref{Candy})), obtain matrices
$\tqbinom{\textbf{Y}^R_i}{\textbf{Y}^I_i}=g^{-1}(\bar{\textbf{y}}_i)$
with $i=1,\cdots,2TN-rank(\mathcal {C})$. Using $\textbf{Y}_i=\textbf{Y}^R_i+j\textbf{Y}^I_i$ as the dispersion matrices in the second group, obtain the resultant 2-group-decodable STBC as
\begin{equation}\label{gp_STBC}
\textbf{X}=s_{1}\textbf{C}_{1}+\sum^{2TN-rank(\mathcal {C})+1}_{l=2}{s_{l}\textbf{Y}_{l-1}}
\end{equation}
where $s_1$ is in the first group, while $s_2$ to $s_{2TN-rank(\mathcal {C})+1}$ are in the second group;

Step 5: Use the constellation rotation technique \cite{Su} to optimize the proposed code. Since the code symbols are divided into mutually orthogonal groups, this constellation rotation can be done group by group.

\subsection{Code Rate} \label{sec_un_coderate}
Since the code rate of unbalanced 2-group-decodable STBC is
$\frac{2TN-rank(\mathcal {C})+1}{2T}$, its upper bound depends on
the lower bound of $rank(\mathcal {C})$. Regarding $rank(\mathcal
{C})$, we have the following theorem:
\begin{theorem}\label{rank_C_prop}
\emph{~\\}
\noindent
\emph{i) When $T\geq N$, i.e., $rank(\textbf{C})=N$, then
$\mathcal {C}$ in (\ref{eq_matrix}) is of full rank and
$rank(\mathcal {C})=N^2$;}

\noindent
\emph{ii) When $T<N$, i.e., $rank(\textbf{C})=T$, then the
lower bound of $rank(\mathcal {C})$ is $2TN-T^2$ and it is reached
when $\textbf{C}$ (after suitable permutations) takes the form of
$\left[\textbf{C}_{sub,T\times T}~\textbf{0}_{T\times
(N-T)}\right]$.}
\end{theorem}

\begin{figure}[!t]
\centering
\includegraphics[width=3.4in]{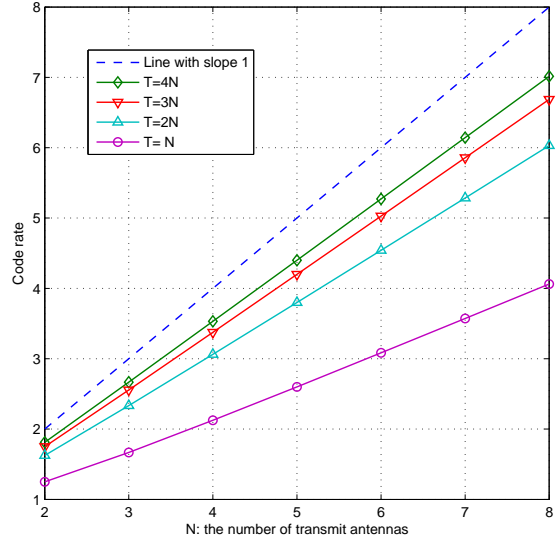}
\caption{Code rates of unbalanced 2-group-decodable STBC for $N$
transmit antennas over $T$ symbol durations. For illustration purpose, $T=N,2N,3N\emph{\emph{ and }}4N$ are
shown.} \label{Rate_un2groupSTBC}
\end{figure}

The proof of Theorem \ref{rank_C_prop} is given in Appendix
\ref{proof_rank_C}. From Theorem \ref{rank_C_prop}, it can be deduced that there are $2TN-rank(\mathcal
{C})=2TN-N^2$ (when $T\geq N$) or $T^2$ (when $T< N$) dispersion matrices in the second
group. Then the following proposition on the maximum code rate of
unbalanced 2-group-decodable STBC can be obtained:
\begin{proposition}
\emph{For an unbalanced 2-group-decodable STBC for $N$ transmit
antennas over $T$ symbol durations, its maximum achievable code rate
is $\frac{2TN-N^2+1}{2T}$ for $T\geq N$, or $\frac{T^2+1}{2T}$ for
$T< N$. For the former, when $T\gg N$, the code rate
$\frac{2TN-N^2+1}{2T}=N-\frac{N^2-1}{2T}$ approaches $N$
asymptotically, i.e., the code approaches full rate.}
\end{proposition}
The code rate variation of the proposed unbalanced
2-group-decodable STBC as a function of $N$ and $T$ is shown in Fig.
\ref{Rate_un2groupSTBC}.

\subsection{ML Decoding Complexity Order} \label{sec_un_complexity}
Clearly, the ML decoding complexity order of the proposed 2-group-decodable STBC is mainly decided

\noindent by that of the larger group. Following \cite{Biglieri}, the ML decoding complexity order $O$ can be shown as:
\begin{equation}\label{DC_order}
O=K\cdot M^{\frac{L_{max}-K+1}{2}}_s=K\cdot 2^{\frac{(L_{max}-K+1)b}{2R}}
\end{equation}
where $L_{max}$ denotes the number of real symbols (need not be orthogonal) in the largest group, $K$ denotes the number of orthogonal real symbols in that group ($K=1$ if the largest group is fully non-orthogonal), $M_s=2^{\frac{b}{R}}$ denotes the size of the signal constellation applied with transmission bit rate $b$ and STBC code rate $R$. For the proposed unbalanced 2-group-decodable STBC, $L_{max}=L_2=2TN-N^2$ (when $T\geq N$) or $T^2$ (when $T< N$), while $R=\frac{2TN-N^2+1}{2T}$ (when $T\geq N$) or $\frac{T^2+1}{2T}$ (when $T< N$).

\subsection{Code Examples }
\subsubsection{2 Transmit Antennas} \label{example_un_nt2}

In this subsection, we present a step-by-step example of the construction of an
unbalanced 2-group-decodable STBC for 4 transmit antennas over 2
symbol durations. This code can be used in 2 ways: (i) to form an unbalanced 2-group-decodable code for 2 transmit antennas; (ii) to construct a balanced 2-group-decodable code for 4 transmit antennas in Section \ref{example_b_nt4}.

Step 1: Pick a 2$\times$4 matrix $\textbf{C}_1$ with full rank (rank 2) as the dispersion matrix in the first group:
\begin{equation}\label{un_2x4_c}
\textbf{C}_1=\left[
\begin{array}{cccccccccc}
    1     &1    & 0 & 0 \\
    1     &-1   & 0 & 0  \\
\end{array}
\right]
\end{equation}
Note that $\textbf{C}_1$ satisfies Theorem \ref{rank_C_prop}(ii), hence it achieves the code rate bound.

Step 2: Based on $\textbf{C}_1$, obtain matrix $\mathcal {C}$ with $rank(\mathcal {C})=2TN-T^2=12$ from (\ref{Candy}):
\begin{equation*}
\mathcal{C}=f(\textbf{C}_1)={\small\left[
\begin{array}{ccccccccccccccccc}
    1  &0  &1  &0  &0  &0  &0  &0  &0  &0  &0  &0  &0  &0  &0  &0\\
    0  &0  &0  &0  &1  &0  &-1 &0  &0  &0  &0  &0  &0  &0  &0  &0\\
    0  &0  &0  &0  &0  &0  &0  &0  &0  &0  &0  &0  &0  &0  &0  &0\\
    0  &0  &0  &0  &0  &0  &0  &0  &0  &0  &0  &0  &0  &0  &0  &0\\
    1  &0  &-1 &0  &1  &0  &1  &0  &0  &0  &0  &0  &0  &0  &0  &0\\
    0  &-1 &0  &1  &0  &1  &0  &1  &0  &0  &0  &0  &0  &0  &0  &0\\
    0  &0  &0  &0  &0  &0  &0  &0  &1  &0  &1  &0  &0  &0  &0  &0\\
    0  &0  &0  &0  &0  &0  &0  &0  &0  &1  &0  &1  &0  &0  &0  &0\\
    0  &0  &0  &0  &0  &0  &0  &0  &0  &0  &0  &0  &1  &0  &1  &0\\
    0  &0  &0  &0  &0  &0  &0  &0  &0  &0  &0  &0  &0  &1  &0  &1\\
    0  &0  &0  &0  &0  &0  &0  &0  &1  &0  &-1 &0  &0  &0  &0  &0\\
    0  &0  &0  &0  &0  &0  &0  &0  &0  &1  &0  &-1 &0  &0  &0  &0\\
    0  &0  &0  &0  &0  &0  &0  &0  &0  &0  &0  &0  &1  &0  &-1 &0\\
    0  &0  &0  &0  &0  &0  &0  &0  &0  &0  &0  &0  &0  &1  &0  &-1\\
    0  &0  &0  &0  &0  &0  &0  &0  &0  &0  &0  &0  &0  &0  &0  &0\\
    0  &0  &0  &0  &0  &0  &0  &0  &0  &0  &0  &0  &0  &0  &0  &0\\
\end{array}
\right]}
\end{equation*}

Step 3: Solve equation (\ref{eq_matrix}) with matrix $\mathcal {C}$, and obtain the solution space of dimension $T^2=4$ as:
\begin{equation*}
\bar{\textbf{y}}=k_1\bar{\textbf{y}}_1+k_2\bar{\textbf{y}}_2+k_3\bar{\textbf{y}}_3+k_4\bar{\textbf{y}}_4
=k_1{\small\left[
\begin{array}{ccccccccccccccccccc}
    -1 \\
    0  \\
    1  \\
    0  \\
    1  \\
    0  \\
    1  \\
    0 \\
    0 \\
    0  \\
    0  \\
    0  \\
    0  \\
    0  \\
    0  \\
    0 \\
\end{array}
\right]}+
 k_2{\small\left[
\begin{array}{ccccccccccccccccccc}
    0 \\
    1  \\
    0  \\
    1  \\
    0  \\
    -1  \\
    0  \\
    1 \\
    0 \\
    0  \\
    0  \\
    0  \\
    0  \\
    0  \\
    0  \\
    0 \\
\end{array}
\right]}+
 k_3{\small\left[
\begin{array}{ccccccccccccccccccc}
    0 \\
    1  \\
    0  \\
    -1  \\
    0  \\
    1  \\
    0  \\
    1 \\
    0 \\
    0  \\
    0  \\
    0  \\
    0  \\
    0  \\
    0  \\
    0 \\
\end{array}
\right]}+ k_4{\small\left[
\begin{array}{ccccccccccccccccccc}
    0 \\
    1  \\
    0  \\
    1  \\
    0  \\
    1  \\
    0  \\
    -1 \\
    0 \\
    0  \\
    0  \\
    0  \\
    0  \\
    0  \\
    0  \\
    0 \\
\end{array}
\right]}
\end{equation*}

Step 4: Under the vector-to-matrix mapping function
$g^{-1}$, obtain $\tqbinom{\textbf{Y}^R_i}{\textbf{Y}^I_i}=g^{-1}(\bar{\textbf{y}}_i)$ with $i=1,2,3\emph{\emph{ and }}4$ as:{\small
\begin{equation*}
\tqbinom{\textbf{Y}^R_1}{\textbf{Y}^I_1}=\left[
\begin{array}{cccccccccc}
    -1    &1    &0     &0\\
    1     &1    &0     &0\\
    0     &0    &0     &0\\
    0     &0    &0     &0\\
\end{array}
\right],~ \tqbinom{\textbf{Y}^R_2}{\textbf{Y}^I_2}=\left[
\begin{array}{cccccccccc}
    0     &0    &0     &0\\
    0     &0    &0     &0\\
    1     &-1   &0     &0\\
    1     &1   &0     &0 \\
\end{array}
\right],~ \tqbinom{\textbf{Y}^R_3}{\textbf{Y}^I_3}=\left[
\begin{array}{cccccccccc}
    0     &0    &0     &0\\
    0     &0   &0     &0 \\
    1     &1    &0     &0\\
    -1    &1    &0     &0\\
\end{array}
\right],~ \tqbinom{\textbf{Y}^R_4}{\textbf{Y}^I_4}=\left[
\begin{array}{cccccccccc}
    0     &0    &0     &0\\
    0     &0    &0     &0\\
    1     &1    &0     &0\\
    1     &-1   &0     &0\\
\end{array}
\right].
\end{equation*}}
So, we have $\textbf{Y}_i=\textbf{Y}^R_i+j\textbf{Y}^I_i$ as:
\begin{equation}\label{un_2x4_y}
\textbf{Y}_1={\small\left[
\begin{array}{cccccccccc}
    -1    &1    &0     &0\\
    1     &1    &0     &0\\
\end{array}
\right]},~ \textbf{Y}_2={\small\left[
\begin{array}{cccccccccc}
    j     &-j   &0     &0\\
    j     &j   &0     &0 \\
\end{array}
\right]},~ \textbf{Y}_3={\small\left[
\begin{array}{cccccccccc}
    j     &j    &0     &0\\
    -j    &j   &0     &0 \\
\end{array}
\right]},~ \textbf{Y}_4={\small\left[
\begin{array}{cccccccccc}
    j     &j    &0     &0\\
    j     &-j    &0     &0\\
\end{array}
\right]}.
\end{equation}
\noindent We emphasize that since (\ref{eq_matrix}) is under-determined, there will be many possible solutions of $\{\bar{\textbf{y}}_i\}$. Typically, we choose the set of $\{\bar{\textbf{y}}_i\}$ leading to
\begin{ventry}{$\bullet$}
\item[$\bullet$] full-rank dispersion matrices, in order to achieve full symbol-wise diversity
gain \cite{Tirkkonen_symbolwisediversity}\cite{Hassibi};
\item[$\bullet$]as many orthogonal dispersion matrices as possible,
in order to achieve a large K in (\ref{DC_order}).
\end{ventry}

Since the dispersion matrices $\textbf{C}_1$ in (\ref{un_2x4_c}) and $\textbf{Y}_1$ to $\textbf{Y}_4$ in (\ref{un_2x4_y}) transmit no information on the third and fourth antennas, so they can be reduced to the followings\footnote{The original $\textbf{C}_1$, $\textbf{Y}_1$ to $\textbf{Y}_4$ will be used in Section \ref{example_b_nt4} to construct a balanced 2-group-decodable code example.} without loss in code rate or diversity:
\begin{equation}\label{un_2x2_cy}
\textbf{C}_1=\left[
\begin{array}{cccccccccc}
    1     &1     \\
    1     &-1    \\
\end{array}
\right];~ \textbf{Y}_1=\left[
\begin{array}{cccccccccc}
    -1    &1   \\
    1     &1   \\
\end{array}
\right],~ \textbf{Y}_2=\left[
\begin{array}{cccccccccc}
    j     &-j   \\
    j     &j    \\
\end{array}
\right],~ \textbf{Y}_3=\left[
\begin{array}{cccccccccc}
    j     &j  \\
    -j    &j   \\
\end{array}
\right],~ \textbf{Y}_4=\left[
\begin{array}{cccccccccc}
    j     &j   \\
    j     &-j  \\
\end{array}
\right].
\end{equation}
Hence an unbalanced 2-group-decodable STBC with the dispersion matrices in (\ref{un_2x2_cy}) for 2 transmit antennas can be obtained as:
\begin{equation}\label{example_un_matrixnt2}
\textbf{X}_{\emph{\emph{un}},2}
=s_{1}\textbf{C}_{1}+\sum^{5}_{l=2}{s_{l}\textbf{Y}_{l-1}}
=\left[
\begin{array}{cccccccccc}
    s_1-s_2+js_3+js_4+js_5     & s_1+s_2-js_3+js_4+js_5 \\
    s_1+s_2+js_3-js_4+js_5     &-s_1+s_2+js_3+js_4-js_5 \\
\end{array}
\right]
\end{equation}
where $s_1$ is in the first group, while $s_2$ to $s_{5}$ are in the second group. Furthermore, $s_2$ to $s_{4}$ are orthogonal, which leads to $L_{max}=4$ and $K=3$ in the decoding complexity order formula (\ref{DC_order}) for this code. $\textbf{X}_{\emph{\emph{un}},2}$ has code rate $R=5/4$, hence, its decoding complexity order calculated following (\ref{DC_order}) is:
\begin{equation}\label{DC_order_un_matrixnt2}
O=K\cdot 2^{\frac{(L_{max}-K+1)b}{2R}}=3\cdot 2^{\frac{4b}{5}}.
\end{equation}

\subsubsection{4 Transmit Antennas}\label{example_un_nt4}

In this subsection, we present the code example of a
2-group-decodable STBC for 4 transmit antennas over 4 symbol
durations.

Step 1: Pick a 4$\times $4 matrix $\textbf{C}_1$ with rank 4 as the
dispersion matrix in the first group:

\begin{equation}\label{un_4x4_c}
\textbf{C}_1=\left[
\begin{array}{cccccccccc}
    1     &0    &0   &0   \\
    0     &1    &0   &0   \\
    0     &0    &1   &0   \\
    0     &0    &0   &1   \\
\end{array}
\right].
\end{equation}

Step 2-4: Since $T=N$, $2TN-N^2=16$
dispersion matrices can be obtained in the second group as:
{\small
\begin{equation*} 
\begin{split}
\textbf{Y}_1&=\left[
\begin{array}{cccccccccc}
    0     &0    &-1   &0   \\
    0     &0    &0   &-1   \\
    1     &0    &0   &0    \\
    0     &1    &0   &0    \\
\end{array}
\right],~~\textbf{Y}_2=\left[
\begin{array}{cccccccccc}
    0     &0    &j   &0   \\
    0     &0    &0   &-j  \\
    j     &0    &0   &0   \\
    0     &-j    &0   &0  \\
\end{array}
\right],~\textbf{Y}_3=\left[
\begin{array}{cccccccccc}
    0     &0    &0   &1   \\
    0     &0    &-1   &0  \\
    0     &1    &0   &0   \\
    -1     &0    &0   &0  \\
\end{array}
\right],~\textbf{Y}_4=\left[
\begin{array}{cccccccccc}
    0     &0    &0   &j   \\
    0     &0    &j   &0   \\
    0     &j    &0   &0   \\
    j     &0    &0   &0   \\
\end{array}
\right],~\\
\end{split}
\end{equation*}
\begin{equation} \label{un_4x4_y}
\begin{split}
\textbf{Y}_5&=\left[
\begin{array}{cccccccccc}
    j     &0    &0   &0   \\
    0     &j    &0   &0   \\
    0     &0    &-j   &0  \\
    0     &0    &0   &-j  \\
\end{array}
\right],~~\textbf{Y}_6=\left[
\begin{array}{cccccccccc}
    0     &1    &0   &0   \\
    -1     &0    &0   &0  \\
    0     &0    &0   &1   \\
    0     &0    &-1   &0  \\
\end{array}
\right],~\textbf{Y}_7=\left[
\begin{array}{cccccccccc}
    0     &j    &0   &0   \\
    j     &0    &0   &0   \\
    0     &0    &0   &-j  \\
    0     &0    &-j   &0  \\
\end{array}
\right],~ \textbf{Y}_8=\left[
\begin{array}{cccccccccc}
    j     &0    &0   &0   \\
    0     &j    &0   &0   \\
    0     &0    &j  &0    \\
    0     &0    &0   &j   \\
\end{array}
\right],\\
\textbf{Y}_{9}&=\left[
\begin{array}{cccccccccc}
    0     &0    &1   &0   \\
    0     &0    &0   &-1  \\
    -1     &0    &0   &0  \\
    0     &1    &0   &0   \\
\end{array}
\right],
~\textbf{Y}_{10}=\left[
\begin{array}{cccccccccc}
    0     &0    &j   &0   \\
    0     &0    &0   &j   \\
    j     &0    &0   &0   \\
    0     &j    &0   &0   \\
\end{array}
\right],\textbf{Y}_{11}=\left[
\begin{array}{cccccccccc}
    j     &0    &0   &0   \\
    0     &-j    &0   &0  \\
    0     &0    &-j   &0  \\
    0     &0    &0   &j   \\
\end{array}
\right],\textbf{Y}_{12}=\left[
\begin{array}{cccccccccc}
    0     &0    &0   &1   \\
    0     &0    &1   &0   \\
    0     &-1    &0   &0  \\
    -1     &0    &0   &0  \\
\end{array}
\right],\\
\textbf{Y}_{13}&=\left[
\begin{array}{cccccccccc}
    0     &1    &0   &0   \\
    -1     &0    &0   &0   \\
    0     &0    &0   &-1    \\
    0     &0    &1   &0   \\
\end{array}
\right],~\textbf{Y}_{14}=\left[
\begin{array}{cccccccccc}
    0     &j    &0   &0   \\
    j     &0    &0   &0   \\
    0     &0    &0   &j    \\
    0     &0    &j   &0   \\
\end{array}
\right],\textbf{Y}_{15}=\left[
\begin{array}{cccccccccc}
    0     &0    &0   &j   \\
    0     &0    &-j   &0   \\
    0     &-j    &0   &0    \\
    j     &0    &0   &0   \\
\end{array}
\right],\textbf{Y}_{16}=\left[
\begin{array}{cccccccccc}
    j     &0    &0   &0   \\
    0     &-j    &0   &0   \\
    0     &0    &j   &0    \\
    0     &0    &0   &-j   \\
\end{array}
\right].
\end{split}
\end{equation}}

The resultant unbalanced 2-group-decodable STBC is:
\begin{equation} \label{example_un_matrixnt4}
\begin{split}
&\textbf{X}_{\emph{\emph{un,4}}}=s_{1}\textbf{C}_{1}+\sum^{17}_{l=2}{s_{l}\textbf{Y}_{l-1}}=\\
&{\small{\small
\left[
\begin{array}{cccccccccc}
 s_1+js_6+js_9+js_{12}+js_{17}         &s_{7}+js_{8}+s_{14}+js_{15}    & -s_{2}+js_{3}+s_{10}+js_{11}   & s_{4}+js_{5}+s_{13}+js_{16}   \\
-s_{7}+js_{8}-s_{14}+js_{15}  &s_1+js_6+js_9-js_{12}-js_{17}           &-s_{4}+js_{5}+s_{13}-js_{16}   & -s_{2}-js_{3}-s_{10}+js_{11}   \\
 s_{2}+js_{3}-s_{10}+js_{11}    & s_{4}+js_{5}-s_{13}-js_{16}   &s_1-js_6+js_9-js_{12}+js_{17}         & s_{7}-js_{8}-s_{14}+js_{15} \\
-s_{4}+js_{5}-s_{13}+js_{16}   &s_{2}-js_{3}+s_{10}+js_{11}     &-s_{7}-js_{8}+s_{14}+js_{15}   &s_1-js_6+js_9+js_{12}-js_{17}   \\
\end{array}
\right]}}
\end{split}
\end{equation}
where $s_1$ is in the first group, while $s_2$ to $s_{17}$ are in the second group. Furthermore, $s_2$ to $s_{6}$ are orthogonal, which leads to $L_{max}=16$ and $K=5$ in the decoding complexity order formula (\ref{DC_order}) for this code. $\textbf{X}_{\emph{\emph{un}},4}$ has code rate $R=17/8$, hence, its decoding complexity order calculated following (\ref{DC_order}) is:
\begin{equation}\label{DC_order_un_matrixnt4}
O=K\cdot 2^{\frac{(L_{max}-K+1)b}{2R}}=5\cdot 2^{\frac{48b}{17}}.
\end{equation}

\subsubsection{3 Time Slots}\label{example_un_t3}
In the 3GPP standardization effort, a 2-antenna STBC that fits into 3 time slots (instead of the typical 2 time slots) are desired due to peculiarity in the existing protocol \cite{3gpp}\cite{ALcode}. Our code construction framework is able to easily obtain a 2-group decodable STBC $\textbf{X}_{\emph{\emph{3gpp}}}$ to meet such atypical specifications, while achieving the maximum rate 3/2 and full symbol-wise diversity.
\begin{equation}\label{example_un_matrixt3}
\textbf{X}_{\emph{\emph{3gpp}}}=s_{1}\textbf{C}_{1}+\sum^{9}_{l=2}{s_{l}\textbf{C}_{l}}={\small{\small
\left[
\begin{array}{cccccccccc}
 s_1+js_2+js_5+js_{6}+js_{9}    &s_{3}+js_{4}+s_{7}+js_{8}    \\
-s_{3}+js_{4}-s_{7}+js_{8}      &s_1-js_2+js_5-js_{6}+js_{9}        \\
 s_{7}+js_{8}                   & s_{6}+js_{9}
\end{array}
\right]}}
\end{equation}
where $s_1$ is in the first group, $s_2$ to $s_{9}$ are in the second group, and
\begin{equation*} 
\begin{split}
\textbf{C}_1&=\left[
\begin{array}{cccccccccc}
    1     &0    \\
    0     &1    \\
    0     &0    \\
\end{array}
\right];\\
\textbf{C}_2&=\left[
\begin{array}{cccccccccc}
    j     &0    \\
    0     &-j    \\
    0     &0    \\
\end{array}
\right],~\textbf{C}_3=\left[
\begin{array}{cccccccccc}
    0     &1    \\
    -1    &0    \\
    0     &0    \\
\end{array}
\right],~\textbf{C}_4=\left[
\begin{array}{cccccccccc}
    0     &j    \\
    j     &0    \\
    0     &0    \\
\end{array}
\right],~\textbf{C}_5=\left[
\begin{array}{cccccccccc}
    j     &0    \\
    0     &j    \\
    0     &0    \\
\end{array}
\right],~\\
\textbf{C}_6&=\left[
\begin{array}{cccccccccc}
    j     &0    \\
    0     &-j    \\
    0     &1    \\
\end{array}
\right],~\textbf{C}_7=\left[
\begin{array}{cccccccccc}
    0     &1    \\
    -1    &0    \\
    1     &0    \\
\end{array}
\right],~\textbf{C}_8=\left[
\begin{array}{cccccccccc}
    0     &j    \\
    j     &0    \\
    j     &0    \\
\end{array}
\right],~\textbf{C}_9=\left[
\begin{array}{cccccccccc}
    j     &0    \\
    0     &j    \\
    0     &j    \\
\end{array}
\right].\\
\end{split}
\end{equation*}
Furthermore, $s_2$ to $s_{4}$ are orthogonal, which leads to $L_{max}=8$ and $K=3$ in the decoding complexity order formula (\ref{DC_order}) for this code. $\textbf{X}_{\emph{\emph{3gpp}}}$ has code rate $R=3/2$, hence, its decoding complexity order calculated following (\ref{DC_order}) is:
\begin{equation} 
O=K\cdot 2^{\frac{(L_{max}-K+1)b}{2R}}=3\cdot 2^{2b}.
\end{equation}

\section{Balanced 2-Group-Decodable STBC}
\subsection{Code Construction}
We now present a method of constructing balanced
2-group-decodable STBC for $N$ transmit antennas over $T$ ($T$ even) symbol
durations from two unbalanced 2-group-decodable STBC.
\begin{proposition}\label{balancedprop}
\emph{Suppose that
$\{\textbf{A}_1;\textbf{A}_2,\cdots,\textbf{A}_{L}\}$ and
$\{\textbf{B}_1;\textbf{B}_2,\cdots,$$\textbf{B}_{L}\}$ are the
dispersion matrices of two unbalanced 2-group-decodable
STBC for $N$ transmit antennas over $\frac{T}{2}$ symbol durations
where $\textbf{A}_1$ satisfies the QOC with
$\textbf{A}_2,\cdots,\textbf{A}_{L}$,~$\textbf{B}_1$ satisfies the
QOC with $\textbf{B}_2,\cdots,\textbf{B}_{L}$,
$\tqbinom{\textbf{A}^R_2}{\textbf{A}^I_2},\cdots,
\tqbinom{\textbf{A}^R_{L}}{\textbf{A}^I_{L}}$ are linearly
independent, and
$\tqbinom{\textbf{B}^R_2}{\textbf{B}^I_2},\cdots,
\tqbinom{\textbf{B}^R_{L}}{\textbf{B}^I_{L}}$ are linearly
independent too. Let
\begin{equation*}
\begin{split}
&\{\textbf{U}_1,\textbf{U}_2,\cdots ,\textbf{U}_{L}\}=
\left\{\left[
\begin{array}{cccc}
    \textbf{A}_2      \\
    \textbf{B}_1 \\
\end{array}
\right],\cdots ,\left[
\begin{array}{cccc}
    \textbf{A}_{L}    \\
    \textbf{B}_1 \\
\end{array}
\right], \left[
\begin{array}{cccc}
    \textbf{A}_i     \\
    -\textbf{B}_1 \\
\end{array}
\right]\right\},
\end{split}
\end{equation*}
\begin{equation*}
\begin{split}
&\{\textbf{V}_1,\textbf{V}_2,\cdots ,\textbf{V}_{L} \}= 
\left\{\left[
\begin{array}{cccc}
    \textbf{A}_1      \\
    \textbf{B}_2 \\
\end{array}
\right],\cdots ,\left[
\begin{array}{cccc}
    \textbf{A}_1      \\
    \textbf{B}_{L} \\
\end{array}
\right], \left[
\begin{array}{cccc}
    -\textbf{A}_1      \\
    \textbf{B}_{k} \\
\end{array}
\right]\right\},
\end{split}
\end{equation*}
where $i,k\in \{2,3,\cdots,L\}$. Then, the matrices
$\textbf{U}_1,\textbf{U}_2,$ $\cdots,\textbf{U}_{L}$ satisfy the
QOC with $\textbf{V}_1,\textbf{V}_2,\cdots,\textbf{V}_{L}$;
$\tqbinom{\textbf{U}^R_1}{\textbf{U}^I_1},$ $\cdots,
\tqbinom{\textbf{U}^R_{L}}{\textbf{U}^I_{L}}$ are linearly
independent, and
$\tqbinom{\textbf{V}^R_1}{\textbf{V}^I_1},\cdots,
\tqbinom{\textbf{V}^R_{L}}{\textbf{V}^I_{L}}$ are linearly
independent too. Note that the
$\{\textbf{A}_1;\textbf{A}_2,\cdots,\textbf{A}_{L}\}$ and
$\{\textbf{B}_1;\textbf{B}_2,\cdots,$$\textbf{B}_{L}\}$ can be the
same or different.}
\end{proposition}
Based on Definition \ref{group_decodable_STBC},
$\{\textbf{U}_1,\textbf{U}_2,\cdots,\textbf{U}_{L}\}$ and
$\{\textbf{V}_1,$ $\textbf{V}_2,\cdots,\textbf{V}_{L}\}$ in
Proposition \ref{balancedprop} can be applied as the dispersion
matrices of a balanced 2-group decodable STBC.

\begin{figure}[!t]
\centering
\includegraphics[width=3.4in]{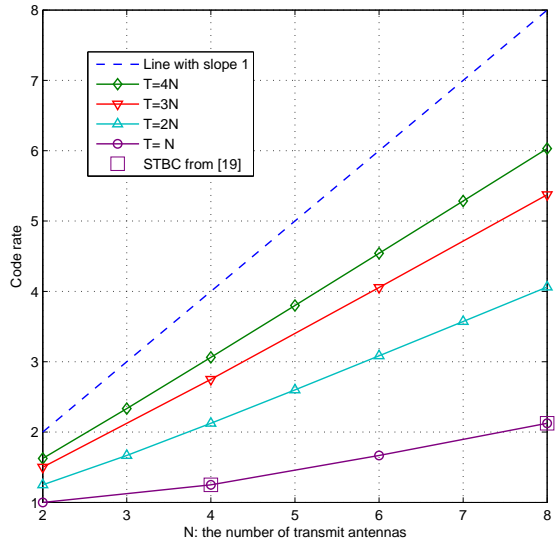}
\caption{Code rates of balanced 2-group-decodable STBC for $N$
transmit antennas over $T$ symbol durations constructed following Proposition
\ref{balancedprop}. For illustration purpose, $T=N,2N,3N\emph{\emph{ and }}4N$ are shown.} \label{Rate_b2groupSTBC}
\end{figure}

\subsection{Code Rate}
\begin{proposition}
\emph{For the balanced 2-group-decodable STBC for $N$ transmit
antennas over $T$ (even) symbol durations constructed following Proposition
\ref{balancedprop}, its code rate can approach $\frac{TN-N^2+1}{T}$ for $T\geq 2N$, or $\frac{T^2+4}{4T}$ for
$T<2N$. For the former, when $T\gg N$, the code rate
$\frac{TN-N^2+1}{T}$ approaches $N$
asymptotically, i.e., the code approaches full rate.}
\end{proposition}
\begin{IEEEproof}
For the dispersion matrices $\textbf{A}_l$ and
$\textbf{B}_l~(l=1,\cdots,L)$ in Proposition \ref{balancedprop}, we
have shown in Section \ref{sec_un_coderate} that the maximum achievable $L$ is
$2\left(\frac{T}{2}\right)N-N^2+1$ (when $\frac{T}{2}\geq N$) or $\left(\frac{T}{2}\right)^2+1$ (when $\frac{T}{2}<N$). Therefore, the balanced 2-group-decodable STBC constructed from Proposition \ref{balancedprop} is of code rate
$\frac{L+L}{2T}=\frac{TN-N^2+1}{T}$ (when $\frac{T}{2}\geq N$) or $\frac{T^2+4}{4T}$ (when $\frac{T}{2}<N$, including $T=N$).

For the former, when $T\gg N$, the code rate $\frac{TN-N^2+1}{T}=N-\frac{N^2-1}{T}$ approaches $N$ asymptotically, i.e., the code approaches full rate.
\end{IEEEproof}

The code rate variation of the proposed balanced 2-group-decodable STBC as a function of $N$ and $T$ is shown in Fig. \ref{Rate_b2groupSTBC}. Note that the 2-group-decodable STBC proposed in \cite{Rajan_2gp} supports $T=N$, $N=2^m~(m\geq2)$ transmit antennas, and code rate $2^{m-2}+\frac{1}{2^m}$. They are indicated as big square markers in Fig. \ref{Rate_b2groupSTBC}. Clearly, our proposed construction is more scalable in code length, transmit antennas number and code rate.

\subsection{Code Example} \label{example_b_nt4}
Following the code construction in Section
\ref{section_un_construction}, another set of dispersion matrices
$\{\textbf{C}'_1;\textbf{Y}'_1,\textbf{Y}'_{2},\textbf{Y}'_3,\textbf{Y}'_{4}\}$
for an unbalanced 2-group-decodable STBC for 4 transmit antennas over 2
symbol durations can be obtained as:
\begin{equation} \label{un_2x4_c2}
\textbf{C}'_1=\left[
\begin{array}{cccccccccc}
    0     &0    & 1     &1    \\
    0     &0    & 1     &-1    \\
\end{array}
\right];
\end{equation}
\begin{equation}\label{un_2x4_y2}
\textbf{Y}'_1={\small\left[
\begin{array}{cccccccccc}
    0     &0    &-1    &1    \\
    0     &0    &1     &1    \\
\end{array}
\right]},~ \textbf{Y}'_2={\small\left[
\begin{array}{cccccccccc}
    0     &0    &j     &-j   \\
    0     &0    &j     &j    \\
\end{array}
\right]},~ \textbf{Y}'_3={\small\left[
\begin{array}{cccccccccc}
    0     &0    &j     &j   \\
    0     &0    &-j    &j   \\
\end{array}
\right]},~ \textbf{Y}'_4={\small\left[
\begin{array}{cccccccccc}
    0     &0    & j     &j    \\
    0     &0    & j     &-j   \\
\end{array}
\right]}.
\end{equation}

Let
$\{\textbf{A}_1;\textbf{A}_2,\textbf{A}_3,\textbf{A}_4,\textbf{A}_5\}$ and $\{\textbf{B}_1;\textbf{B}_2,\textbf{B}_3,\textbf{B}_4,\textbf{B}_5\}$ in Proposition \ref{balancedprop} be the $\{\textbf{C}_1;\textbf{Y}_1,\textbf{Y}_{2},\textbf{Y}_3,\textbf{Y}_{4}\}$
in (\ref{un_2x4_c})(\ref{un_2x4_y}) and $\{\textbf{C}'_1;\textbf{Y}'_1,\textbf{Y}'_{2},\textbf{Y}'_3,\textbf{Y}'_{4}\}$ in (\ref{un_2x4_c2})(\ref{un_2x4_y2}), then the dispersion matrices for a balanced 2-group-decodable STBC obtained are:
\begin{equation*} 
\begin{split}
&\textbf{U}_1={\small\left[
\begin{array}{cccccccccc}
    1     &1   &0     &0    \\
    1     &-1  &0     &0     \\
    0     &0   &-1    &1    \\
    0     &0   &1    &1    \\
\end{array}
\right]},~\textbf{U}_2={\small\left[
\begin{array}{cccccccccc}
    1     &1   &0     &0    \\
    1     &-1  &0     &0     \\
    0     &0   &j    &-j    \\
    0     &0   &j    &j    \\
\end{array}
\right]},~\textbf{U}_3={\small\left[
\begin{array}{cccccccccc}
    1     &1   &0     &0    \\
    1     &-1  &0     &0     \\
    0     &0   &j    &j    \\
    0     &0   &-j    &j    \\
\end{array}
\right]},~ \textbf{U}_4={\small\left[
\begin{array}{cccccccccc}
    1     &1   &0     &0    \\
    1     &-1  &0     &0     \\
    0     &0   &j    &j    \\
    0     &0   &j    &-j    \\
\end{array}
\right]},\\
\end{split}
\end{equation*}
\begin{equation}\label{b_4x4_cy}
\begin{split}
&\textbf{U}_5={\small\left[
\begin{array}{cccccccccc}
    -1    &-1   &0     &0    \\
    -1    &1  &0     &0     \\
    0     &0   &-1    &1    \\
    0     &0   &1    &1    \\
\end{array}
\right]};~\textbf{V}_1={\small\left[
\begin{array}{cccccccccc}
    -1     &1   &0     &0    \\
    1     &1  &0     &0     \\
    0     &0   &1    &1    \\
    0     &0   &1    &-1    \\
\end{array}
\right]},~\textbf{V}_2={\small\left[
\begin{array}{cccccccccc}
    j     &-j   &0     &0    \\
    j     &j  &0     &0     \\
    0     &0   &1    &1    \\
    0     &0   &1    &-1    \\
\end{array}
\right]},~ \textbf{V}_3={\small\left[
\begin{array}{cccccccccc}
    j     &j   &0     &0    \\
    -j     &j  &0     &0     \\
    0     &0   &1    &1    \\
    0     &0   &1    &-1    \\
\end{array}
\right]},\\&\textbf{V}_4={\small\left[
\begin{array}{cccccccccc}
    j     &j   &0     &0    \\
    j     &-j  &0     &0     \\
    0     &0   &1    &1    \\
    0     &0   &1    &-1    \\
\end{array}
\right]},~~~\textbf{V}_5={\small\left[
\begin{array}{cccccccccc}
    -1     &1   &0     &0    \\
    1     &1  &0     &0     \\
    0     &0   &-1    &-1    \\
    0     &0   &-1    &1    \\
\end{array}
\right]}.
\end{split}
\end{equation}

The resultant balanced 2-group-decodable STBC for 4 transmit antennas over 4
symbol durations is:
\begin{equation}\label{example_b_matrixnt2}
\begin{split}
&\textbf{X}_{\emph{\emph{b,4}}}=\sum^{5}_{l=1}{s_{l}\textbf{U}_{l}}+\sum^{10}_{l=6}{s_{l}\textbf{V}_{l-5}}=
\\&{\small\left[\begin{array}{cccccccccccccccc}
    s_1+s_2+s_3+s_4-s_5+js_6+js_7+js_8-s_9-s_{10}   ~~~~ s_1+s_2+s_3+s_4-s_5+js_6-js_7+js_8+s_9+s_{10}  ~~~~~0   ~~~0   \\
    s_1+s_2+s_3+s_4-s_5+js_6+js_7-js_8+s_9+s_{10}   ~~~ -s_1-s_2-s_3-s_4+s_5-js_6+js_7+js_8+s_9+s_{10} ~~~0   ~~~0   \\
    0   ~~~0 ~~~js_1+js_2+js_3-s_4-s_5+s_6+s_7+s_8+s_9-s_{10}   ~~~~~js_1-js_2+js_3+s_4+s_5+s_6+s_7+s_8+s_9-s_{10}    \\
    0   ~~~0 ~~~js_1+js_2-js_3+s_4+s_5+s_6+s_7+s_8+s_9-s_{10}   ~~-js_1+js_2+js_3+s_4+s_5-s_6-s_7-s_8-s_9+s_{10}   \\
\end{array}
\right]}
\end{split}
\end{equation}
where $s_1$ to $s_5$ are in the first group, while $s_{6}$ to $s_{10}$ are in the second group. For this code, $L_{max}=5$, $K=1$ and $R=5/4$, hence, its decoding complexity order calculated following (\ref{DC_order}) is:
\begin{equation}\label{DC_order_b_matrixnt2}
O=K\cdot 2^{\frac{(L_{max}-K+1)b}{2R}}= 2^{2b}.
\end{equation}

\section{Simulations and Discussions}
In this section, we investigate the BER performance and ML decoding complexity order of the 2-group-decodable STBC examples shown earlier. In all simulations, the MIMO channel is assumed to be quasi-static Rayleigh fading in the sense that the channel coefficients do not change
during one codeword transmission, and the channel state information
is perfectly known at the receiver.

\subsection{Unbalanced 2-Group-Decodable STBC}
\subsubsection{2 Transmit Antennas}
In this subsection, we compare the unbalanced
2-group-decodable STBC $\textbf{X}_{\emph{\emph{un,2}}}$ in (\ref{example_un_matrixnt2}) with Alamouti code \cite{Alamouti}, BLAST \cite{Foschini} and Golden code \cite{Belfiore} in a 2$\times$2
MIMO system with 4 bits per channel use. Due to the different code rates, Alamouti code, BLAST
and Golden code are simulated with 16-QAM, 4-QAM and 4-QAM,
respectively. On the other hand, $\textbf{X}_{\emph{\emph{un,2}}}$ is of code
rate $5/4$. We let one real symbol be drawn from 4-PAM, and other 4 real symbols
(viewed as 2 complex symbols) be drawn from 8-QAM, then the bit rate
of $\textbf{X}_{\emph{\emph{un,2}}}$ is 4 bits per channel use.

The parameters of these codes are compared in Table \ref{table_comp_2gp2nt}, including the decoding complexity order following (\ref{DC_order}). Table \ref{table_comp_2gp2nt} shows that the proposed code has much lower decoding complexity order than Golden code due to group-decodable code structure, and higher decoding complexity order than Alamouti code due to higher code rate. For example, with $b=4$ bits per channel use, the decoding complexities of Golden code, the proposed code and Alamouti code are in decreasing order of $2^8,~ 3\cdot 2^{3}(\emph{\emph{approximate}})\emph{\emph{ and }}2^2$.
\begin{center}
\begin{threeparttable}[!b]
\tabcolsep 5mm \caption{Comparison of Space-Time Codes in A 2$\times$2 MIMO System with $b$ Bits/Channel Use.} \label{table_comp_2gp2nt}
\newcommand{\rb}[1]{\raisebox{1.8ex}[0pt]{#1}}
\newcommand{\rbb}[1]{\raisebox{1.0ex}[0pt]{#1}}
{\small{\small{\small\begin{tabular}{| c |c | c | c | c |c |} \hline
&  &  &  & \multicolumn{2} {|c|}{Complexity order: $O$}
\\\cline{5-6}  & \rb{Code length: $T$} &
\rb{Code rate: $R$} & \rb{Group size: $L_{max}$} & $b$ & $b=4$
\\ \hline\hline
{\rule[-1mm]{0mm}{6mm}}Alamouti code\cite{Alamouti} &2    & 1   & 1
& $2^{\frac{b}{2}}$ & $2^2$
\\ \hline
{\rule[-1mm]{0mm}{6mm}}BLAST\cite{Foschini}        & 1   & 2   & 4 &
$2^b$ & $2^4$
\\ \hline
{\rule[-1mm]{0mm}{6mm}}Golden code\cite{Belfiore}  & 2    & 2  & 8 &
$2^{2b}$ & $2^8$
\\ \hline
{\rule[-1mm]{0mm}{6mm}}$\textbf{X}_{\emph{\emph{un,2}}}$ proposed in
(\ref{example_un_matrixnt2}) & 4 & 5/4 &4 & $3\cdot
2^{\frac{4b}{5}}$ & $\approx 3\cdot 2^{3}$
\\ \hline
\end{tabular}
}}}
\end{threeparttable}
\end{center}
\vspace{0.04in}

\begin{figure}[!t]
\centering
\includegraphics[width=4.3in]{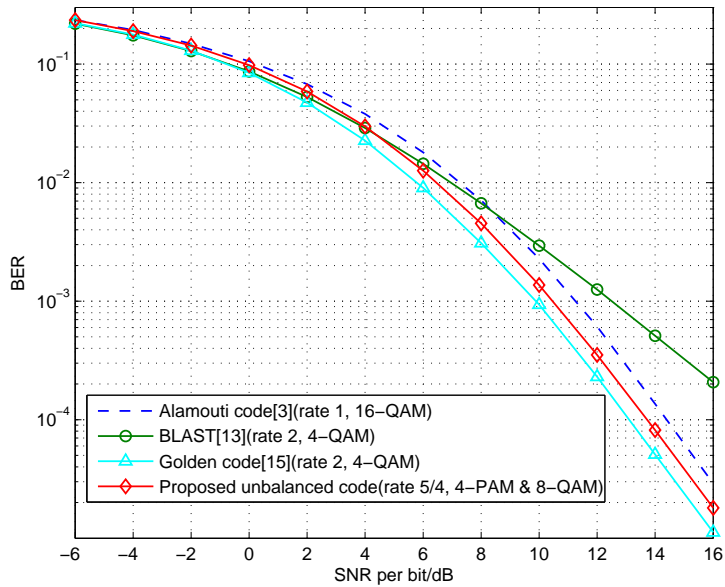}
\caption{BER performances in a 2$\times$2 MIMO system with 4 bits
per channel use.}
\label{ber_nt2_4bps}
\end{figure}

We plot the BER curves of these codes in Fig. \ref{ber_nt2_4bps}. To achieve full diversity, constellation rotations for $\textbf{X}_{\emph{\emph{un,2}}}$ are obtained by computer search\footnote{Optimized constellation rotation angles are 0 for $s_4$ (drawn from 4-PAM), $0.0735\pi$ for $s_1$ and $s_5$ (drawn from 8-QAM), $0$ for $s_2$ and $s_3$ (drawn from 8-QAM).}. Fig. \ref{ber_nt2_4bps} shows that the proposed $\textbf{X}_{\emph{\emph{un,2}}}$ can
achieve full diversity (same BER slope as Alamouti code and Golden code). The BER curve of $\textbf{X}_{\emph{\emph{un,2}}}$ lies between those of Golden code and Alamouti code, which is in accordance with their code rates.

\subsubsection{4 Transmit Antennas}
We compare the proposed unbalanced 2-group-decodable STBC with
orthogonal STBC (OSTBC) \cite{Ganesan_ostbc}, quasi-orthogonal STBC
(QOSTBC) \cite{Tirkkonen,Jafarkhani,Yuen_book}, perfect code
\cite{Oggier} and PS-SR code \cite{Rajan4x2} in a 4$\times$2 MIMO
system with about 4 bits per channel use. The OSTBC, QOSTBC, perfect
code and PS-SR code are simulated with 32-QAM (3.75 bits per channel
use), 16-QAM, 16-QAM (3.64 bits per channel use), 4-QAM and 4-QAM,
respectively. To achieve code rate 2, we simulate the unbalanced
2-group-decodable STBC $\textbf{X}_{\emph{\emph{un,4}}}$ in
(\ref{example_un_matrixnt4}) with $s_{17}$ removed. Then the bit
rate of $\textbf{X}_{\emph{\emph{un,4}}}$ with 4-QAM is 4 bits per
channel use.

The parameters of these codes are listed in Table \ref{table_comp_2gp4nt}. It shows that the proposed code has lower ML decoding complexity order than the perfect code due to group-decodable code structure, and higher decoding complexity order than the OSTBC and QOSTBC due to higher code rate. Due to fast decoding code structure, the PS-SR code has a lower decoding complexity that that of the proposed. For example,
\begin{center}
\begin{threeparttable}[!b]
\tabcolsep 3.4mm \caption{Comparison of Space-Time Codes in A 4$\times$2 MIMO System with $b$ Bits/Channel Use.}\label{table_comp_2gp4nt}
\newcommand{\rb}[1]{\raisebox{1.8ex}[0pt]{#1}}
\newcommand{\rbb}[1]{\raisebox{1.0ex}[0pt]{#1}}
{\small{\small{\small\begin{tabular}{| c |c | c | c | c |c |} \hline
&  &  &  & \multicolumn{2} {|c|}{Complexity order: $O$}
\\\cline{5-6}  & \rb{Code length: $T$} &
\rb{Code rate: $R$} & \rb{Group size: $L_{max}$} & $b$ & $b=4$
\\ \hline\hline
{\rule[-1mm]{0mm}{6mm}}OSTBC\cite{Ganesan_ostbc} &4    & 3/4   &1 &
$2^{\frac{2b}{3}}$ & $\approx 2^{3}$
\\ \hline
{\rule[-1mm]{0mm}{6mm}}QOSTBC\cite{Tirkkonen,Jafarkhani,Yuen_book} &
4 & 1   & 2 & $2^b$ & $2^{4}$
\\ \hline
{\rule[-1mm]{0mm}{6mm}}Perfect code\cite{Oggier}  & 4    & 2  & 16 &
$2^{4b}$ & $2^{16}$
\\ \hline
{\rule[-1mm]{0mm}{6mm}}PS-SR code\cite{Rajan4x2}  & 4    & 2  & 16 &
$8\cdot2^{\frac{9b}{4}}$ & $8\cdot 2^{9}$
\\ \hline
{\rule[-1mm]{0mm}{6mm}}$\textbf{X}_{\emph{\emph{un,4}}}$ in
(\ref{example_un_matrixnt4}) with $s_{17}$ removed$^a$ & 4 & 2& 15 & $5\cdot
2^{\frac{11b}{4}}$ & $5\cdot 2^{11}$
\\ \hline
\end{tabular}
\begin{tablenotes}
\item[$^a$] With $s_{17}$ removed from $\textbf{X}_{\emph{\emph{un,4}}}$ in (\ref{example_un_matrixnt4}), the $R$ and $L_{max}$ in (\ref{DC_order_un_matrixnt4}) are updated as $R=2$ and $L_{max}=15$.
\end{tablenotes}}}}
\end{threeparttable}
\end{center}
\vspace{0.02in}

\noindent with $b=4$ bits per channel use, the decoding complexities
of perfect code, the proposed code, PS-SR code, QOSTBC, and OSTBC
are in decreasing order of $2^{16},~ 5\cdot 2^{11},~8\cdot
2^{9},~2^{4},\emph{\emph{ and }}~2^{3}$ (approximate).

\begin{figure}[!t]
\centering
\includegraphics[width=4.3in]{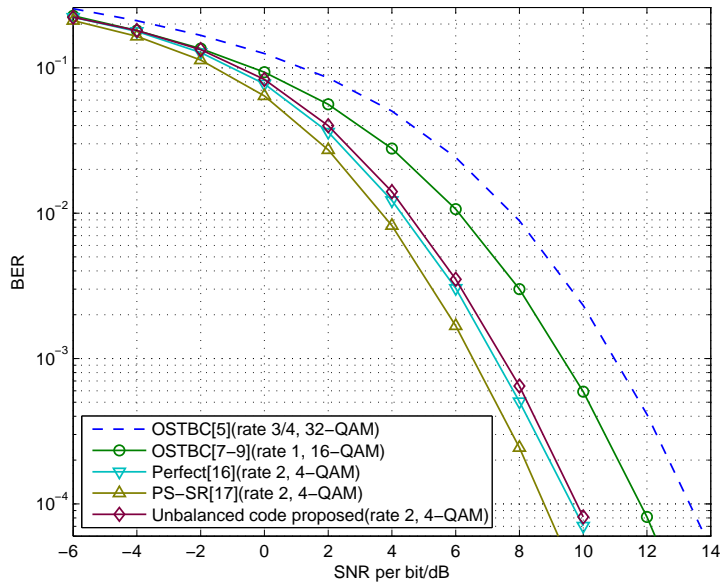}
\caption{BER performances in a 4$\times$2 MIMO system with about 4 bits per channel use.}
\label{ber_nt4_4bps}
\end{figure}

We plot the BER curves in Fig. \ref{ber_nt4_4bps}, where the optimum
constellation rotation proposed in \cite{Su} is applied for QOSTBC
and the constellation rotations for
$\textbf{X}_{\emph{\emph{un,4}}}$ are obtained by computer
search\footnote{Optimized constellation rotation angles are $0$ for
$s_{1}$ and $s_{5}$, $0.1413\pi$ for $s_{2}$ and $s_{6}$,
$0.1413\pi$ for $s_{3}$ and $s_{4}$, $0.1538\pi$ for $s_{7}$ and
$s_{8}$, $0.2493\pi$ for $s_{9}$ and $s_{10}$, $0.1691\pi$ for
$s_{11}$ and $s_{13}$, $0.1044\pi$ for $s_{12}$ and $s_{16}$,
$0.2140\pi$ for $s_{14}$ and $s_{15}$.}. From Fig.
\ref{ber_nt4_4bps}, we can see that the proposed
$\textbf{X}_{\emph{\emph{un,4}}}$ has the same full diversity gain
as the perfect code and the PS-SR code (the PS-SR code has the best BER
performance), and performs much better than OSTBC and QOSTBC due to
higher code rate.

\subsection{Balanced 2-Group-Decodable STBC}
In this subsection, we compare the proposed balanced 2-group-decodable STBC $\textbf{X}_{\emph{\emph{b,4}}}$ in (\ref{example_b_matrixnt2}) with the 2-group-decodable STBC presented in \cite{Rajan_2gp} in a 4$\times$2 MIMO system with 2.5 bits per channel use. Since their code rates are 5/4, they will be simulated with 4-QAM.

\begin{figure}[!t]
\centering
\includegraphics[width=4.3in]{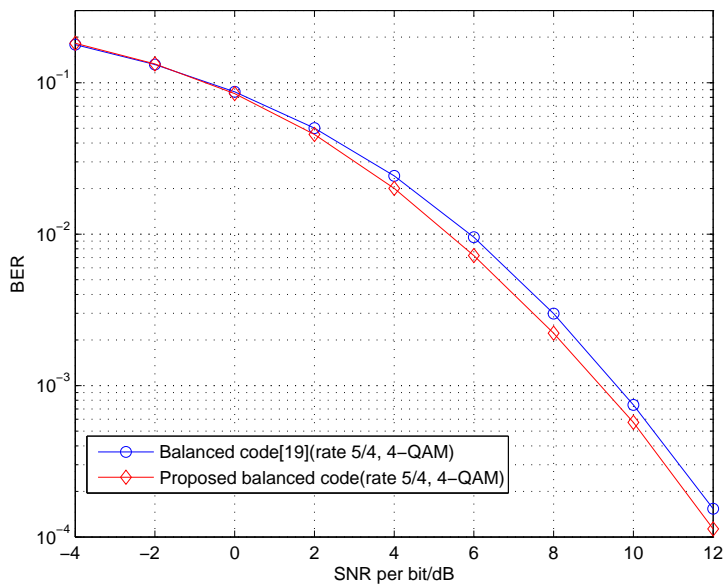}
\caption{BER performances in a 4$\times$2 MIMO system with 2.5 bits per channel use.}
\label{ber_nt4_25bps}
\end{figure}

Both codes have the same decoding complexity order. We plot their BER curves in Fig.
\ref{ber_nt4_25bps}, where the constellation rotations for $\textbf{X}_{\emph{\emph{b,4}}}$ are obtained by computer search\footnote{Optimized constellation angles are $0.1538\pi$ for $s_1$ and $s_3$ (similarly $s_6$ and $s_8$), $0.4625\pi$ for $s_2$ and $s_5$ (similarly $s_7$ and $s_{10}$), 0 for $s_4$ and $s_9$.}. Such constellation rotation optimization are feasible because the information symbols are group-decodable and hence can be optimized separately. From Fig. \ref{ber_nt4_25bps}, we can see that both codes achieve full diversity
gain, and the proposed code has a small 0.3 dB coding gain over the code in \cite{Rajan_2gp} probably because our constellation rotation angles are slightly more optimal.

\subsection{3-Time-Slot STBC}
In this subsection, we compare the proposed 3-time-slot STBC $\textbf{X}_{\emph{\emph{3gpp}}}$ in (\ref{example_un_matrixt3}) with the other 3-time-slot STBC $\textbf{X}_{\emph{\emph{AL}}}$ presented in \cite{ALcode} in a 2$\times$2 MIMO system with 3 bits per channel use. As the code rates of $\textbf{X}_{\emph{\emph{3gpp}}}$ and $\textbf{X}_{\emph{\emph{AL}}}$ are 1.5 and 1, they are applied with 4-QAM and 8-PSK, respectively.

\begin{figure}[!t]
\centering
\includegraphics[width=4.3in]{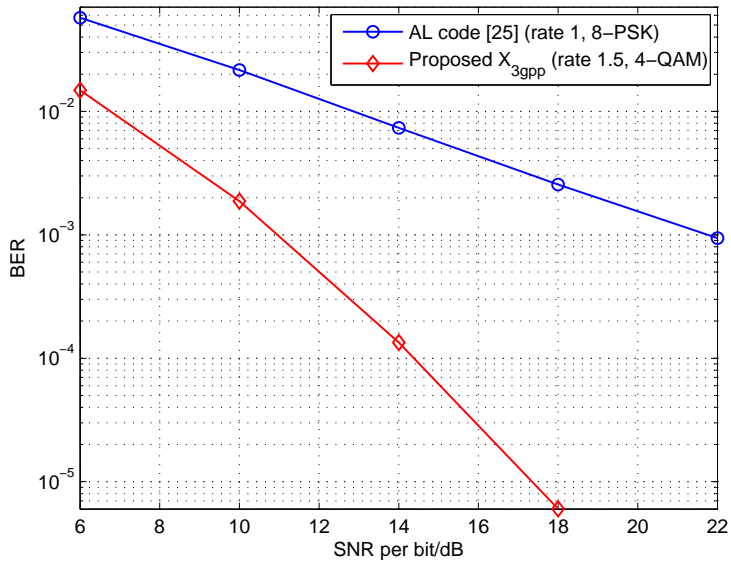}
\caption{BER performances of 3-time-slot codes in a 2$\times$2 MIMO system with 3 bits per channel use.}
\label{ber_t3_3bps}
\end{figure}

We plot their BER curves in Fig.
\ref{ber_t3_3bps}, where the constellation rotations for $\textbf{X}_{\emph{\emph{3gpp}}}$ are obtained by computer search\footnote{Optimized constellation angles are $0$ for $s_1$, $0.0875\pi$ for $s_2$ and $s_6$, $0.0875\pi$ for $s_3$ and $s_7$, $0.05\pi$ for $s_4$ and $s_8$ and $0.1625\pi$ for $s_5$ and $s_9$.}. From Fig. \ref{ber_t3_3bps}, we can see that the proposed $\textbf{X}_{\emph{\emph{3gpp}}}$ achieves a much better performance than the $\textbf{X}_{\emph{\emph{AL}}}$ \cite{ALcode} due to higher diversity gain.

\section{Conclusion}
In this paper, we first derive unbalanced 2-group-decodable
high-rate STBC for $N$ transmit antennas over $T$ symbol durations
with code rates upper-bounded by $\frac{2TN-N^2+1}{2T}$ for $T\geq N$, or
$\frac{T^2+1}{2T}$ for $T<N$, then use them to systematically
construct balanced 2-group-decodable high-rate STBC with code rates
$\frac{TN-N^2+1}{T}$ for $T\geq 2N$, or $\frac{T^2+4}{4T}$ for
$T<2N$. The proposed high-rate STBC are able to achieve full
symbol-wise diversity, and their code rates increase almost linearly
with the transmit antenna number $N$ and approach $N$ asymptotically
when $T\gg N$. Performance studies show that with constellation
rotation optimization, the proposed 2-group-decodable STBC can
achieve the same full diversity as the algebraic STBC, and much
better BER performance than the (quasi-)orthogonal STBC. The
proposed code is very scalable in code length, transmit antenna
number and code rate. Its constellation
rotation optimization is also easier to perform because its
symbols are group-orthogonal and hence can be optimized separately.

\appendices
\section{}
\label{proofindependence} We employ proof by contradiction. 
\renewcommand{\theequation}{\thesection.\arabic{equation}}

 (Necessary condition) Assume that
$\tqbinom{\textbf{C}^R_{i_1}}{\textbf{C}^I_{i_1}},~
\tqbinom{\textbf{C}^R_{i_2}}{\textbf{C}^I_{i_2}},\cdots,$ $
\tqbinom{\textbf{C}^R_{i_{L_i}}}{\textbf{C}^I_{i_{L_i}}}$ are not
linearly independent, i.e.,
$\mathscr{C}_{i_1},\mathscr{C}_{i_2},\cdots,$
$\mathscr{C}_{i_{L_i}}$ in (\ref{r_Hs}) are not linearly
independent. Then there exists $ \alpha_{i_1}\mathscr{C}_{i_1}+
\alpha_{i_2}\mathscr{C}_{i_2} +\cdots
+\alpha_{i_{L_i}}\mathscr{C}_{i_{L_i}}=\textbf{0}$ where not all the
scalars $\alpha_{i_1},\alpha_{i_2},\cdots,\alpha_{i_{L_i}}$ are
zero. Since $\textbf{h}_{i}=\mathscr{C}_i\bar{\textbf{h}}$, we have
\begin{equation}
\begin{split}
\alpha_{i_1}\textbf{h}_{i_1}+\alpha_{i_2}\textbf{h}_{i_2}+\cdots+\alpha_{i_{L_i}}\textbf{h}_{i_{L_i}}
&=\alpha_{i_1}\mathscr{C}_{i_1}\bar{\textbf{h}}+
\alpha_{i_2}\mathscr{C}_{i_2}\bar{\textbf{h}} +\cdots
+\alpha_{i_{L_i}}\mathscr{C}_{i_{L_i}}\bar{\textbf{h}}
\\&=(\alpha_{i_1}\mathscr{C}_{i_1}+ \alpha_{i_2}\mathscr{C}_{i_2}
+\cdots +\alpha_{i_{L_i}}\mathscr{C}_{i_{L_i}}
)\bar{\textbf{h}}\\&=\textbf{0}
\end{split}
\end{equation}
In other words, the assumed premise on $\textbf{h}_{i_1},\textbf{h}_{i_2},\cdots,$ $\textbf{h}_{i_{L_i}}$ is violated.
Therefore, the necessary condition is proved.

(Sufficient condition) Assume that
$\textbf{h}_{i_1},\textbf{h}_{i_2},\cdots,\textbf{h}_{i_{L_i}}$ are
not linearly independent, i.e., $ \alpha_{i_1}\textbf{h}_{i_1}+
\alpha_{i_2}\textbf{h}_{i_2} +\cdots
+\alpha_{i_{L_i}}\textbf{h}_{i_{L_i}}=\textbf{0}$ where not all the
scalars $\alpha_{i_1},\alpha_{i_2},\cdots,\alpha_{i_{L_i}}$ are
zero. We can obtain that:
\begin{equation}
\begin{split}
\textbf{0}&=\alpha_{i_1}\textbf{h}_{i_1}+
\alpha_{i_2}\textbf{h}_{i_2} +\cdots
+\alpha_{i_{L_i}}\textbf{h}_{i_{L_i}}\\&=(\alpha_{i_1}\mathscr{C}_{i_1}+
\alpha_{i_2}\mathscr{C}_{i_2} +\cdots
+\alpha_{i_{L_i}}\mathscr{C}_{i_{L_i}}
)\bar{\textbf{h}}\\&=\mathscr{C}\bar{\textbf{h}}
\end{split}
\end{equation}
where $\bar{\textbf{h}}$ is of size $2NM\times 1$. Since
$\bar{\textbf{h}}$ is the channel coefficient vector with
independent entries, we have $dim(\{\bar{\textbf{h}}\})=2NM$. Then,
$rank(\mathscr{C})$ must be 0. In other words,
$\mathscr{C}=\textbf{0}$. Therefore,
$\mathscr{C}_{i_1},\mathscr{C}_{i_2},\cdots,\mathscr{C}_{i_{L_i}}$
are linearly dependent, i.e.,
$\tqbinom{\textbf{C}^R_{i_1}}{\textbf{C}^I_{i_1}},~
\tqbinom{\textbf{C}^R_{i_2}}{\textbf{C}^I_{i_2}},\cdots,
\tqbinom{\textbf{C}^R_{i_{L_i}}}{\textbf{C}^I_{i_{L_i}}}$ are not
linearly independent. Hence, the sufficient condition is proved.

Combining the two conclusions, Theorem \ref{th_independence}
is proved.

\section{}
\label{proof_rank_C}

\begin{figure*}
\centering
\includegraphics[width=7.2in]{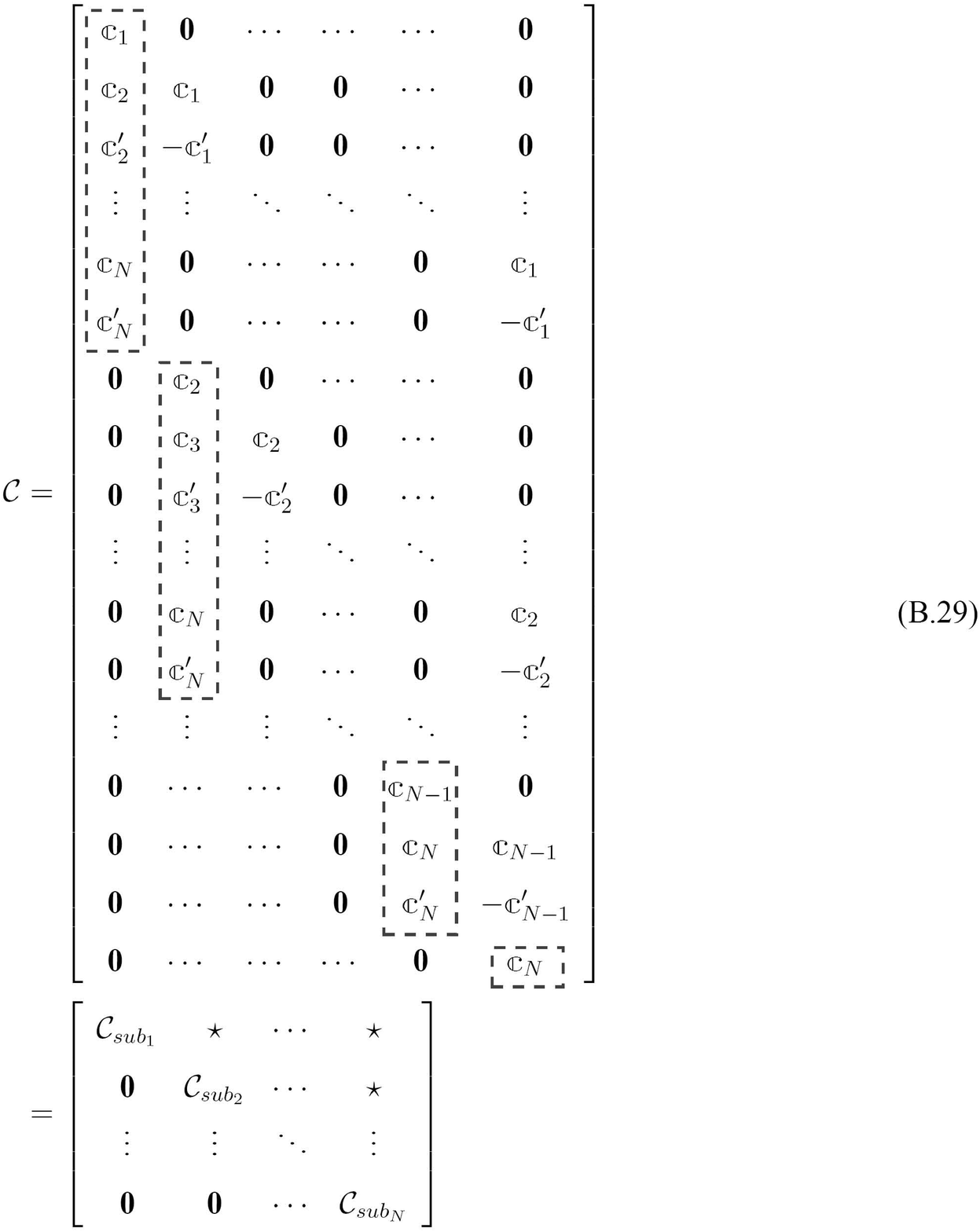}
\end{figure*}

\begin{figure*}
\centering
\includegraphics[width=7.2in]{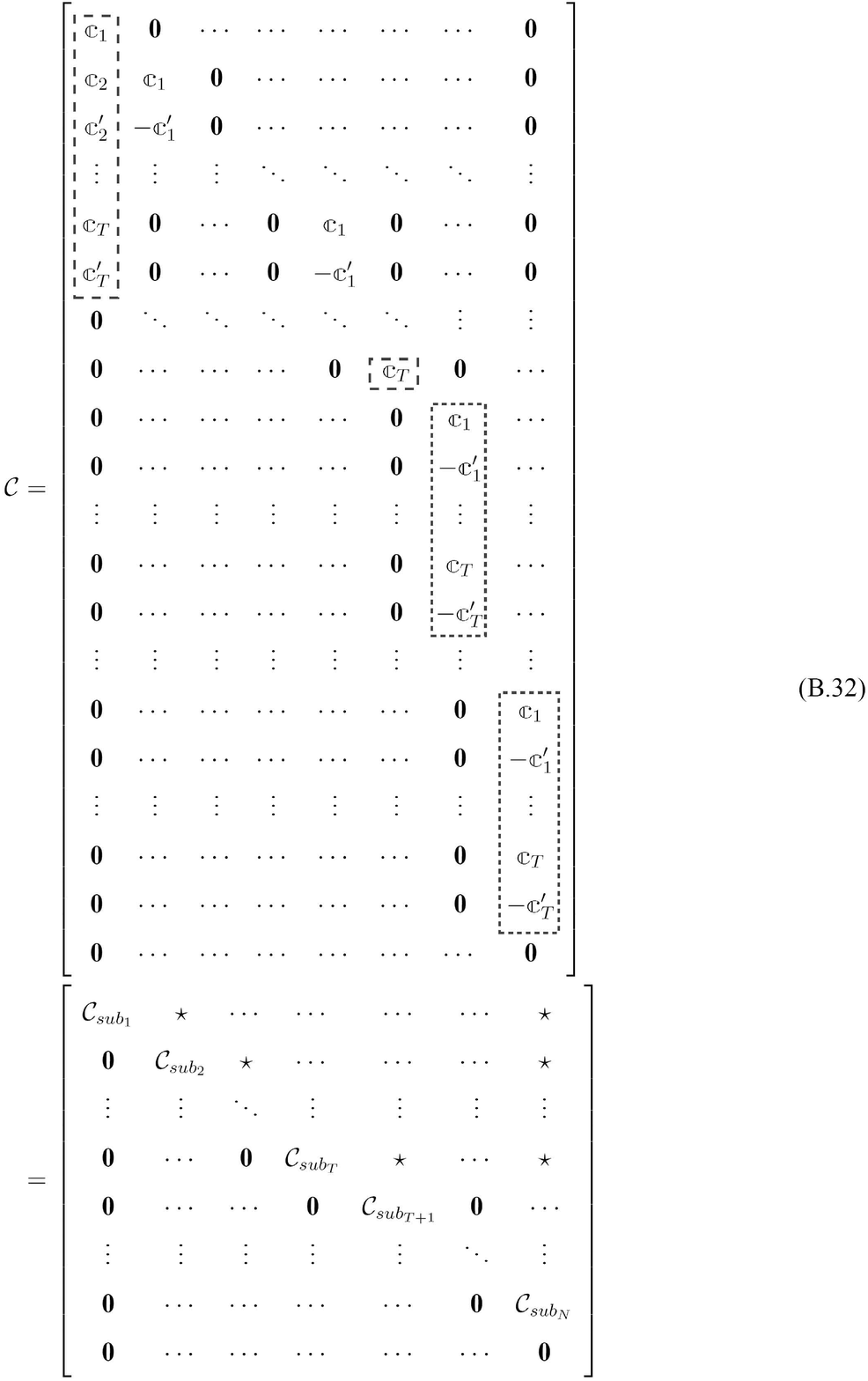}
\end{figure*}

i) When $T\geq N$, after some row/column permutations,
$\mathcal {C}$ can be rewritten as (B.29) where $\mathcal
{C}_{sub_i}=[\mathbbm{c}_i~\mathbbm{c}_{i+1}~\mathbbm{c}'_{i+1}~\cdots$
$\mathbbm{c}_N~\mathbbm{c}'_N ]^T(i=1,2,\cdots ,N)$ are highlighted
in dashed boxes and $\star$ stand for the other
elements in $\mathcal {C}$. In fact,
$rank(\mathcal{C}_{sub_i})=2(N-i)+1$ because
$\mathbbm{c}_1,\mathbbm{c}_{2},\mathbbm{c}'_{2},\cdots,\mathbbm{c}_N,\mathbbm{c}'_N$
are linearly-independent row vectors when $rank(\textbf{C})=N$. This
will be proved below using proof by contradiction.

Recall that
$\mathbbm{c}_n=[c^R_{1n}~c^I_{1n}~c^R_{2n}~c^I_{2n}~\cdots~
c^R_{Tn}~c^I_{Tn}]$ and $\mathbbm{c}'_n=[c^I_{1n}~-c^R_{1n}$
$~c^I_{2n}~-c^R_{2n}~\cdots~c^I_{Tn}~-c^R_{Tn}]$. Let
$\textbf{C}=[\textbf{c}_1~\textbf{c}_2,\cdots~\textbf{c}_N]$, then
$\mathbbm{c}_i=e(\textbf{c}_i)~(i=1,2,\cdots ,N)$ where $e$ is a
complex-vector-to-real-vector mapping function and
$\mathbbm{c}'_i=e(-j\textbf{c}_i)$. Suppose that
$\mathbbm{c}_1,\mathbbm{c}_{2},\mathbbm{c}'_{2},\cdots,\mathbbm{c}_N,\mathbbm{c}'_N$
are linearly dependent, then, since
$\mathbbm{c}_1,\mathbbm{c}_{2},\mathbbm{c}'_{2},\cdots,\mathbbm{c}_N,\mathbbm{c}'_N$
are real, there is \setcounter{equation}{29}
\begin{equation}\label{real_dependent}
\alpha_{11}\mathbbm{c}_1+
\alpha_{21}\mathbbm{c}_{2}+\alpha_{22}\mathbbm{c}'_{2}+\cdots
+\alpha_{N1}\mathbbm{c}_{N}+\alpha_{N2}\mathbbm{c}'_{N}=\textbf{0}
\end{equation}
where not all the real scalars
$\alpha_{11},\alpha_{21},\alpha_{22},\cdots,\alpha_{N1},\alpha_{N2}$
are zero. Therefore, under the inverse function for $e$,
(\ref{real_dependent}) can be presented as \setcounter{equation}{30}
\begin{equation}
\alpha_{11}\textbf{c}_1+
(\alpha_{21}-j\alpha_{22})\textbf{c}_2+\cdots
+(\alpha_{N1}-j\alpha_{N2})\textbf{c}_N=\textbf{0}
\end{equation}
Since not all the values
$\alpha_{11},\alpha_{21}-j\alpha_{22},\alpha_{N1}-j\alpha_{N2}$ are
zero, $\textbf{c}_1,\textbf{c}_2,\cdots,\textbf{c}_N$ are linearly
dependent and $rank(\textbf{C})<N$, which is contrary to the
original premise. Therefore,
$\mathbbm{c}_1,\mathbbm{c}_{2},\mathbbm{c}'_{2},\cdots,\mathbbm{c}_N,\mathbbm{c}'_N$
are linearly independent, $rank(\mathcal{C}_{sub_i})=2(N-i)+1$ and
$rank(\mathcal{C})=\sum_{i=1}^{N}{rank(\mathcal{C}_{sub_i})}=N^2$.

ii) When $T<N$, $\textbf{C}_{N\times T}$ can be written as
$\left[\textbf{C}_{sub}~\textbf{0}\right]$ where
$rank(\textbf{C})=rank(\textbf{C}_{sub})=T$. After some row/column
permutation, $\mathcal {C}$ can be rewritten as (B.32) where
$\mathcal
{C}_{sub_i}=[\mathbbm{c}_i~\mathbbm{c}_{i+1}~\mathbbm{c}'_{i+1}~
\cdots~\mathbbm{c}_N~\mathbbm{c}'_N ]^T$ are highlighted in
\emph{dashed} boxes with $i = 1, 2, \cdots ,T$ and $\mathcal
{C}_{sub_i}=[\mathbbm{c}_{1}~-\mathbbm{c}'_{1}~\cdots~
\mathbbm{c}_{T}$ $-\mathbbm{c}'_{T}]^T$ are highlighted in
\emph{dotted} boxes with $i=T+1,\cdots ,N$. As stated above, as
$rank(\textbf{C}_{sub})=T$, it can be proved by contradiction that
$\mathbbm{c}_1,\mathbbm{c}_{2},\mathbbm{c}'_{2},
\cdots,\mathbbm{c}_T,\mathbbm{c}'_T $ are linearly independent, and
$\mathbbm{c}_{1},-\mathbbm{c}'_{1},\cdots,\mathbbm{c}_{T},-\mathbbm{c}'_{T}$
are linearly independent too. Then, $rank(\mathcal
{C}_{sub_i})=2(T-i)+1$ for $i=1,2,\cdots ,T$ and $rank(\mathcal
{C}_{sub_i})=2T$ for $i=T+1,\cdots ,N$. Therefore,
$rank(\mathcal{C})=\sum_{i=1}^{N}{rank(\mathcal{C}_{sub_i})}=2TN-T^2$.

If $\textbf{C}$ is not of the form
$\left[\textbf{C}_{sub}~\textbf{0}\right]$, it is easy to prove that
$rank(\mathcal{C})\geq 2TN-T^2$. Due to space limitation, the proof
is omitted here.

\end{document}